\documentclass{article}

\usepackage{PRIMEarxiv}

\usepackage[utf8]{inputenc} % allow utf-8 input
\usepackage[T1]{fontenc}    % use 8-bit T1 fonts
\usepackage{hyperref}       % hyperlinks
\usepackage{url}            % simple URL typesetting
\usepackage{booktabs}       % professional-quality tables
\usepackage{amsfonts}       % blackboard math symbols
\usepackage{nicefrac}       % compact symbols for 1/2, etc.
\usepackage{microtype}      % microtypography
\usepackage{lipsum}
\usepackage{fancyhdr}       % header
\usepackage{graphicx}       % graphics
\usepackage{amsmath}
\usepackage{amsfonts}
\graphicspath{{media/}}     % organize your images and other figures under media/ folder

\title{Breaking the Degrees-of-Freedom Limit of Holographic MIMO Communications: \\A 3-D Antenna Array Topology}

\author{Shuai S. A. Yuan $^{1}$, Jie Wu $^{2}$, Hongjing Xu $^{3}$, Tengjiao Wang $^{3}$, Da Li $^{1}$, Xiaoming Chen $^{4}$,
		\\ \textbf{Chongwen Huang $^{1}$, Sheng Sun $^{5}$, Shilie Zheng $^{1}$, Xianmin Zhang $^{1}$, Er-Ping Li $^{1}$ and Wei E. I. Sha $^{1,}$*}\\ \\
$^{1}$ College of Information Science and Electronic Engineering, Zhejiang University, Hangzhou 310027, China.\\
$^{2}$ Anhui Province Key Laboratory of Simulation and Design for Electronic Information System, \\ Hefei Normal University, Hefei 230601, China.\\
$^{3}$ Wireless Network Research Department, Huawei Technologies, Shanghai, 201206, China.\\
$^{4}$ School of Information and Communications Engineering, Xi'an Jiaotong University, Xi'an 710049, China.\\
$^{5}$ School of Electronic Science and Engineering, University of Electronic Science and Technology of China, \\ Chengdu 611731, China.\\ \\
\texttt{Correspondence: weisha@zju.edu.cn}}

\begin{document}
\maketitle
\begin{abstract}
The performance of holographic multiple-input multiple-output (MIMO) communications, employing two-dimensional (2-D) planar antenna arrays, is typically compromised by finite degrees-of-freedom (DOF) stemming from limited array size. The DOF constraint becomes significant when the element spacing approaches approximately half a wavelength, thereby restricting the overall performance of MIMO systems. To break this inherent limitation, we propose a novel three-dimensional (3-D) antenna array that strategically explores the untapped vertical dimension. We investigate the performance of MIMO systems utilizing 3-D arrays across different multi-path scenarios, encompassing Rayleigh channels with varying angular spreads and the 3rd generation partnership project (3GPP) channels. We subsequently showcase the advantages of these 3-D arrays over their 2-D counterparts with the same aperture sizes. As a proof of concept, a practical dipole-based 3-D array, facilitated by an electromagnetic band-gap (EBG) reflecting surface, is conceived, constructed, and evaluated. The experimental results align closely with full-wave simulations, and channel simulations substantiate that the DOF and capacity constraints of traditional holographic MIMO systems can be surpassed by adopting such a 3-D array configuration.
\end{abstract}

% keywords can be removed

%%%%%%%%%%%%%%%%%%%%%%%%%%%%%%%%%%%%%%%%
\section{Introduction}
The multiple-input-multiple-output (MIMO) technology, pivotal in wireless communications, has demonstrated substantial capacity enhancement across diverse scenarios, marking considerable success over the years \cite{telatar1999capacity, tse2005fundamentals, paulraj2004overview}. In the pursuit of augmenting MIMO communication performance, several technologies have been developed. Some representative technologies include massive MIMO communications \cite{TL2014, bjornson2019massive,Merouane2013,Wang2023}, decoupling and decorrelation of MIMO arrays \cite{Wu2017, zhang2019mutual,chen2021simultaneous,wang2021improvement}, reconfigurable intelligent surface (RIS) aided MIMO communications \cite{cui2020,Chongwen2019}, and holographic MIMO communications \cite{wei2022multi, An2023,Wei2023, ShuaiOJAP, Zhang2023,Demi2022, Gong2023, gong2023holographic, gong2023transmit}. Holographic MIMO communications, an advanced iteration of RIS, present promising avenues for research in the realm of highly adaptable antennas by skillfully harnessing electromagnetic (EM) waves \cite{Pizzo2020,an2023tutorial1,an2023tutorial2,an2023tutorial3}. Notably, a holographic MIMO array can comprise a large amount of antennas within a compact aperture, and it has been substantiated to possess numerous advantages \cite{Huang2020}. 

However, the primary challenge associated with holographic MIMO communications is the finite degrees-of-freedom (DOF) constrained by array size \cite{Thomas2020}. Given that a holographic MIMO array integrates a substantial number of antennas with sub-wavelength inter-element spacing within a confined area, a pronounced correlation emerges among the antennas, resulting in performance degradation. To compensate for such an impaired performance, an increase in array size is expected. {Unfortunately, in practice, the array sizes at the base stations or vehicles are strictly confined due to wind drag, city planning, product requirements, etc.} Hence, exceeding the DOF limit is imperative in further enhancing the performance of holographic MIMO communications.

In the pursuit of surpassing the DOF limit, it is necessary to make clear the fundamental limits of holographic MIMO communications. This analysis has been undertaken from dual perspectives—both information and electromagnetic (EM)—to provide a comprehensive understanding of the inherent limitations \cite{MD2019, Shuai2021, Dai2023, Han2023, Jeon2018,franceschetti_2017,gustafsson2023degrees,Dardari2021,Li2023, bai2024information}. In a typical multi-path environment, the performance of a MIMO system depends on both the power gain and the DOF gain \cite{tse2005fundamentals}. The power gain is known as the beamforming gain, which is constricted by the aperture size of the array \cite{balanis2015antenna}. One famous argument is Hannan's limit \cite{Hannan1964}, providing the closed-form expression of the radiation efficiency limit of antenna elements when more and more elements are placed into a constrained planar aperture. In addition to the power gain, the DOF gain usually attracts more attention when dealing with MIMO communications. 
\begin{table}
	\centering
	%{center}
	\caption{{Parameters and variables used in this paper}}
	\setlength{\tabcolsep}{5pt}
	\begin{tabular}{|c|c|}
		\hline Names &   Definitions \\
		\hline$\rho_{mn}$ & \begin{tabular}{l} 
			Correlation between the $n$th and $m$th antennas
		\end{tabular} \\
		
		\hline$\Psi$ & \begin{tabular}{l} 
			Diversity measure of the MIMO array
		\end{tabular} \\
		\hline$N_t$, $N_r$ & \begin{tabular}{l} 
			Number of the transmitting and receiving antennas
		\end{tabular} \\
		\hline$h$ & \begin{tabular}{l} 
			Height difference in 3-D antenna array
		\end{tabular} \\
		\hline$\boldsymbol{\Phi}$ & \begin{tabular}{l} 
			Correlation matrix
		\end{tabular} \\
		\hline$e^{emb}$& \begin{tabular}{l} 
			Embedded radiation efficiency
		\end{tabular} \\
		\hline$\Omega$& \begin{tabular}{l} 
			Solid angle
		\end{tabular} \\
		\hline$\lambda_0$, $k_0$& \begin{tabular}{l} 
			Free-space wavelength and wavenumber
		\end{tabular} \\
		\hline$k_x$, $k_y$& \begin{tabular}{l} 
			Wavenumber components along the $x$ and $y$ directions
		\end{tabular} \\
		\hline$\mathbf{k}_{\Omega}$& \begin{tabular}{l} 
			Wave vector along the $\Omega$ direction
		\end{tabular} \\
		\hline$ E(\Omega)$& \begin{tabular}{l} 
			Far-field electric-field pattern
		\end{tabular} \\
		\hline$\gamma$& \begin{tabular}{l} 
			Total signal-to-noise ratio
		\end{tabular} \\
		\hline$\sigma$& \begin{tabular}{l} 
			Eigenvalue of the correlation matrix
		\end{tabular} \\
		\hline$S_{mn}$& \begin{tabular}{l} 
			S parameter between the port $n$ and port $m$
		\end{tabular} \\
		\hline
	\end{tabular}
	\label{tab1}
	%\end{center}
\end{table}
The EM DOF of a MIMO system refers to the effective rank (the number of significant eigenvalues) of its correlation matrix \cite{tse2005fundamentals}, the number of scattering channels \cite{Bucci1989}, or the number of available spatial EM modes \cite{ShuaiPra}, which characterizes the spatial-multiplexing performance of the MIMO system. The DOF of a MIMO system has been discussed based on various models, including the general scattering method \cite{Bucci1989, MD2019, Mats2021,Wallace2008,signal2005space}, the Green's function method \cite{Shuai2021, ShuaiOJAP, Shuai2022, Shen2023, piestun2000electromagnetic}, the intuitive methods \cite{ShuaiPra, miller2000communicating}, the plane-wave expansion model \cite{Pizzo2020, wei2022multi}, etc. These models are essentially equivalent but using different expansion bases of EM waves. {As a summary, the EM DOF limit of an aperture-constrained MIMO system is considered to be only dependent on the surface area of the transmitting/receiving space \cite{piestun2000electromagnetic,MD2019}, or the average scattering cross-section\cite{gustafsson2023degrees,franceschetti_2017}. }

Therefore, one possible route to violate the DOF limit of holographic MIMO communications is to build a three-dimensional (3-D) array, which explores the additional volume available in the vertical dimension, and thus extra DOF. {The performance of a 3-D array has been examined in line-of-sight (LOS) scenarios \cite{hu2018beyond, Song2015}, revealing marginal benefits at far field, but has not yet been explored in rich-scattering environments. Also, some similar concepts can be found in using the scatterers over the MIMO array for realizing the de-correlation (thus improving the DOF and capacity) \cite{wang2021improvement, chen2021simultaneous}, and employing the stacked intelligent metasurfaces for advanced computation and signal processing tasks \cite{an2023stacked,an2023stacked2}.} {As far as our knowledge extends, neither the concept of breaking the DOF limit by using the 3-D array nor the corresponding performance analyses have been discussed before.} In this work, a novel 3-D array topology is proposed for breaking the DOF limit of holographic MIMO systems built with two-dimensional (2-D) arrays. The contributions of this paper can mainly be attributed to the following three aspects: 
\begin{figure}[ht!]
	\centering
	\includegraphics[width=3.2in]{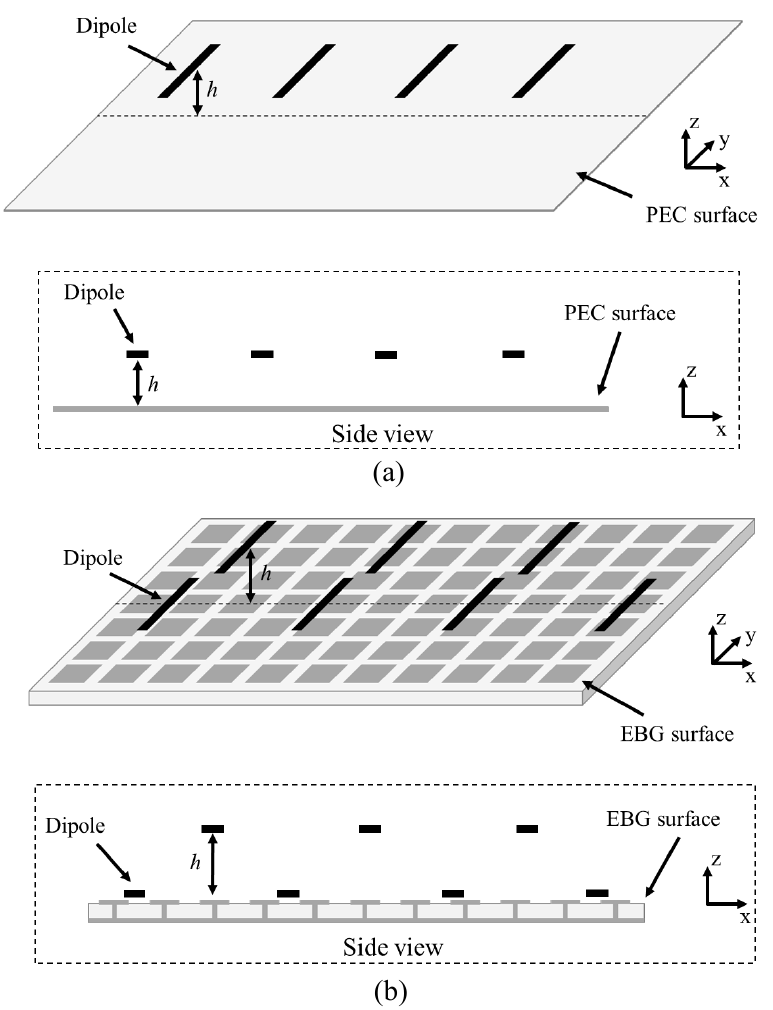}
	\caption{Diagrams of the traditional 2-D antenna array and the proposed 3-D antenna array, where a one-row array is set up for demonstration. (a) Traditional 2-D antenna array over a PEC surface, where the heights of antennas are the same. (b) Proposed 3-D antenna array over an EBG surface and a height difference is introduced between the nearby antennas.}%=d_x\times N=120
	\label{1}
\end{figure}%=======figure==========%
\begin{itemize} 
	\item The theoretical analyses of the MIMO performances of 3-D arrays based on the 3-D Clarke and Kronecker models; for example, comparing the performances of the 3-D and 2-D arrays with the same aperture size at different height differences, element spacings, and angular spreads. 
	\item A practical 3-D antenna array enabled by an electromagnetic band-gap (EBG) reflecting surface is designed, fabricated, and measured as a proof of concept.
	\item Assessments of the MIMO performances of the practical 3-D array are conducted in both Rayleigh channels with varying angular spreads and 3rd generation partnership project (3GPP) channels.
\end{itemize}
%The findings reveal that the 3-D array holds significant promise as a candidate for augmenting MIMO performance in the landscape of future wireless communications.

The rest of this paper is organized as follows. In Section II, the methodology of the 3-D array, and the tools for evaluating MIMO performance, i.e., the 3-D Clarke and Kronecker models, are introduced. Next, the theoretical analyses are carried out in Section III based on the two models. After that, the design, simulation, and experimental validation of a practical 3-D array, together with the discussion of its MIMO performances in Rayleigh and 3GPP channels, are presented in Section IV. Finally, some remarks and conclusions are given in Section V. Table \ref{tab1} summarizes a list of the frequently used parameters and variables in this paper.

%{$Notation$: Fonts $a$, $\mathbf{a}$, and $\mathbf{A}$ represent scalars, vectors, and matrices, respectively. $\mathbf{A}^T$, $\mathbf{A}^H$,  and $||\mathbf{A}||_F$ denote the transpose, Hermitian (conjugate transpose), and Frobenius norm of $\mathbf{A}$. $\mathbf{A}_{i, j}$ represents $\mathbf{A}$’s $(i, j)$-th element, $^*$ is the conjugate operator, $\circ$ denotes the entry-wise product, and tr(·) gives the trace of a matrix. $\mathbf{I}_n$ (with $n \ge 2$) is the $n\times n$ identity matrix. Table I summarizes a list of the frequently used parameters and variables in this paper.}

%
\section{Methodology}
The diagrams of the traditional 2-D array topology and the proposed 3-D array topology are depicted in Fig. \ref{1}. Without loss of generality, a one-row array is set up along the $x$-axis to reduce the complexities of analysis and design. Rather than using the traditional 2-D planar array with a perfect electric conductor (PEC) reflector in Fig. \ref{1}(a), a height difference is introduced between the nearby antennas for realizing the 3-D array in Fig. \ref{1}(b). The lower antennas are close to the EBG surface while the upper antennas maintain a suitable height difference, which can be enabled by a well-designed EBG surface. In the 3-D array, the spatial differences between the array elements are not only along the transverse ($x$ and $y$) directions, but also along the longitudinal  ($z$) direction. 

\begin{figure}[ht!]
	\centering
	\includegraphics[width=2.6in]{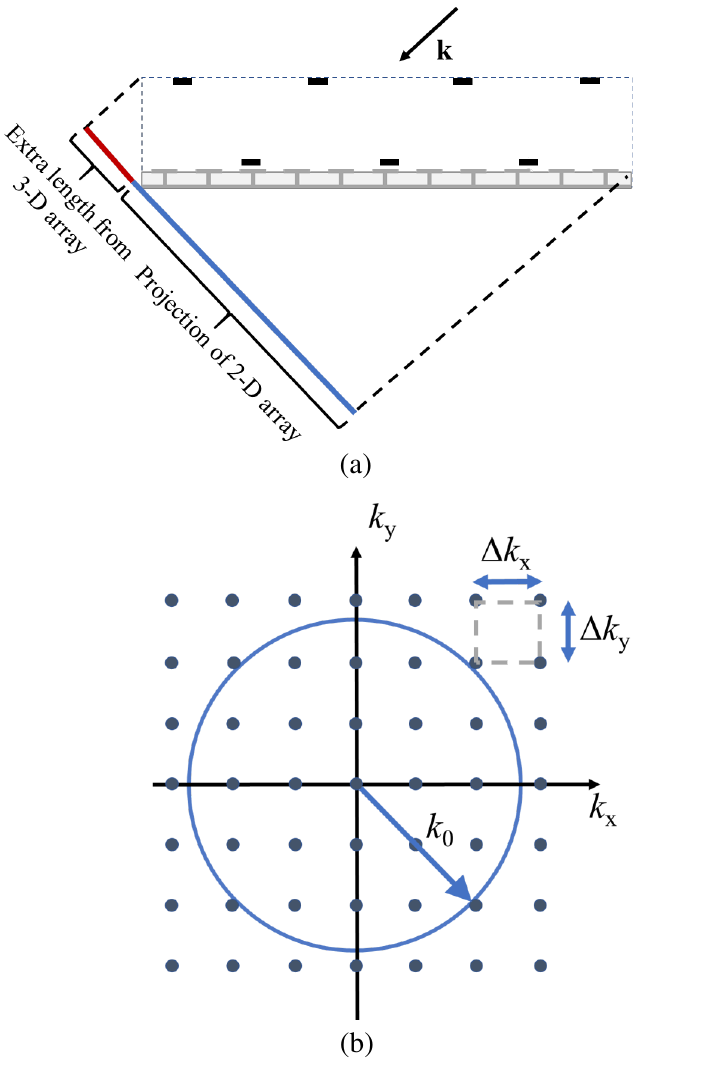}
	\caption{{Intuitive explanations of the benefits from a 3-D antenna array. (a) Under the incidence of a large-angle plane wave with the direction of $\mathbf{k}$, the 3-D array would have a larger projection area compared to the 2-D array. (b) In the angular spectrum analysis of the DOF limit, the blue circle with the radius of $k_0$ represents the available angular spectrum resource, $\Delta k_x$ and $\Delta k_y$ denote the resolutions along the $x$ and $y$ directions. The 3-D array would have better resolutions compared to the 2-D array at large angles.}}%=d_x\times N=120
	\label{2}
\end{figure}

Intuitively, the distance between the nearby antennas becomes larger because of the introduced height difference, and thus the spatial correlation between them will become lower (i.e., a larger DOF), as the spatial correlation between two antennas is usually negatively correlated to the distance between them. From another perspective, the area of the oblique projection of a 3-D array on the transverse plane is always larger than the area of the aperture of a 2-D array, as illustrated in Fig. \ref{2}(a). The above explanations indicate that the additional volume along the vertical dimension can be utilized for realizing an equivalently larger 2-D array. 

{Particularly, a remark regarding the far-field angular spectrum analysis is given, as shown in Fig. \ref{2}(b). The angular domain for propagating waves is defined as $k_x^2+k_y^2 \le k_0^2$, where $k_0$ is the free-space wavenumber, $k_x$ and $k_y$ are the wavenumber components along the $x$ and $y$ directions. Assuming a planar surface with lengths $L_x$ and $L_y$ along the $x$ and $y$ directions, the resolutions of the beams generated by this surface along $k_x$ and $k_y$ at the far-field angular domain would be $\Delta k_x = 2\pi/L_x$ and $\Delta k_y = 2\pi/L_y$. Then, the DOF limit can be easily estimated by dividing the circular area (resource of angular domain $\pi k_0^2$) by the area of a dotted rectangle (resolution of a beam in angular domain $\Delta k_x\Delta k_y$) \cite{ShuaiPra}. In this analysis, the resolutions $\Delta k_x = 2\pi/L_x$ and $\Delta k_y = 2\pi/L_y$ are the same for the beams towards different directions. In fact, the resolutions for beams towards large angles would be worse, since the resolutions would depend on the projection areas of the array along different directions, as shown in Fig. \ref{2}(a). Hence, the resource of angular domain is fixed for any far-field communication, while the 3-D array topology is expected to improve the resolutions of beams towards large angles, thus bring more DOF for MIMO communications.} To further explore the properties of 3-D array, including the benefits and potential drawbacks, the 3-D Clarke and Kronecker models are first used for characterizing their MIMO performances, as discussed below.
%=======figure==========%
\subsection{3-D Clarke model}
Based on the 3-D Clarke model \cite{clarke1968statistical, clarke1997}, the correlations between antennas in a Rayleigh fading environment can be analytically modeled for the theoretical analysis of MIMO performance. {In the Clarke model, the antennas are simplified as point receivers, and the incident waves are modeled as uniformly distributed far-field plane waves. The angular spread of the incident waves can be modified for approximately characterizing different scattering environments, as the spatial correlation is dominated by the angular spread, but not the specific distribution of the incident waves \cite{Andersen2002}. An illustration of angular spread is depicted in Fig. \ref{3}, and the mean incident angle is taken along the broadside of the array.} {The total signal received by one antenna can be regarded as the integral of these received plane waves along solid angles (with a sufficient number of plane waves). Therefore, one can easily formulate the correlation between the two antennas at the positions $\mathbf{r}_m$ and $\mathbf{r}_n$ by}
\begin{equation}
\rho_{nm}=\frac{1}{{N}} \int_{\Omega} \exp (j\mathbf{k}_\Omega \mathbf{r}_n- j\mathbf{k}_\Omega \mathbf{r}_m )\mathrm{d}\Omega,
\end{equation}
where $N$ is the number of arriving plane waves, $\Omega$ represents the solid angles of the plane waves within the angular spread, and $\mathbf{k}_\Omega$ denotes the wave vector along the $\Omega$ direction. Then, the correlation matrix $\boldsymbol{\Phi}$ of the array can be obtained for evaluating the MIMO performance, i.e., the DOF and capacity. 

{DOF indicates the spatial multiplexing performance of a MIMO communication system; in (7.12) of reference \cite{tse2005fundamentals}, DOF is defined as the rank of correlation matrix, i.e., the number of significant eigenvalues. Physically, these eigenvalues are related to the EM eigenmodes generated from the array \cite{miller2019waves}. 
	\begin{figure}[ht!]
		\centering
		\includegraphics[width=2.8in]{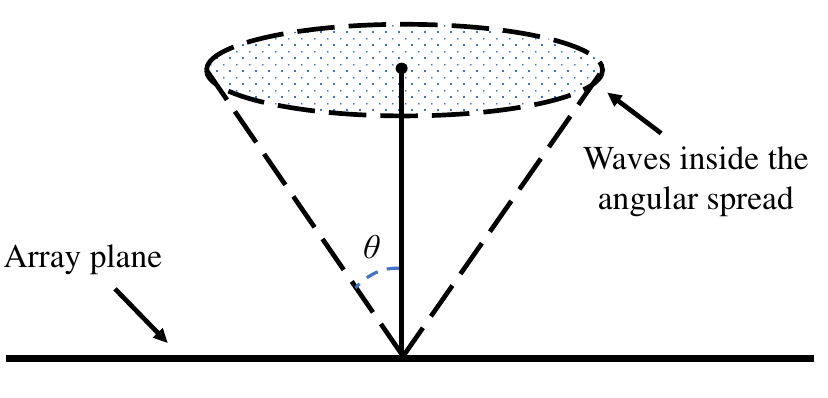}
		\caption{{Illustration of the angular spread, where the plane waves are uniformly distributed in an angular range characterized by $\theta$.}}%=d_x\times N=120
		\label{3}
	\end{figure}%=======figure==========%	
	The DOF limit of a MIMO system is generally dependent on the size of array but not the antenna number, where the eigenvalues corresponding to the saturated antennas are too small to contribute to DOF \cite{ShuaiPra}. Other than DOF, the diversity measure \footnote{{Antenna diversity is a scheme that sends multiple copies of a signal to improve the reliability of data reception. The diversity measure here works as an indicator for the performance of MIMO spatial multiplexing, but can also be regarded as a measurement of antenna diversity, since they are closely related. For example, if the antennas are highly correlated, they are equivalent to much fewer independent antennas when using antenna diversity.}} is also frequently used for characterizing the spatial multiplexing performance \cite{NJ2005, Verdu2002, Shuai2021, zhang2019mutual}, also called effective DOF. Diversity measure $\Psi$, which can approximately represent the equivalent number of isolated antennas of a MIMO array, can be calculated by \cite{NJ2005}}
\begin{equation}
\Psi\left(\boldsymbol{\Phi}\right)=\left(\frac{\operatorname{tr}\left(\boldsymbol{\Phi}\right)}{\left\|\boldsymbol{\Phi}\right\|_{F}}\right)^{2}=\frac{\left(\sum_{i} \sigma_{i}\right)^{2}}{\sum_{i} \sigma_{i}^{2}},
\end{equation}
where $\operatorname{tr}(\cdot)$ represents the trace operator and $\sigma_{i}$ is the $i$th eigenvalue of the correlation matrix $\boldsymbol{\Phi}$. From (2), we can find that both DOF and diversity can well characterize the limit of spatial multiplexing, since no more eigenvalues can contribute to the increase of diversity measure after reaching the limit. The concept of DOF is more physically intuitive compared to the diversity. However, the diversity would be more helpful to capacity estimation, since the number of equivalent isolated antennas is quite straightforward when characterizing a practical MIMO array. Both the DOF and diversity can be used for well capturing the performance limit of spatial multiplexing, i.e., the limited number of significant eigenvalues of correlation matrix. {In this work, we choose to use the diversity measure due to its engineering convenience, and the diversity measure used here should also be distinguished from the antenna diversity for stabilizing wireless link.} To avoid any confusion, we have clearly clarified the definitions of the DOF and diversity, since the definitions of these terms may be discrepant in different works or societies.

Furthermore, we evaluate the capacity of MIMO systems under the vertical-bell-labs-space-time (V-BLAST) architecture \cite{tse2005fundamentals}, where the transmitting side is ideal (the antennas are uncorrelated and their efficiencies are equal to 1). The channel matrix is unknown to the transmitters, and the transmitting power is equally allocated. This architecture allows us to focus on the characteristics of the array at the receiving side. The ergodic capacity incorporating antenna effects can be written as \cite{xiaoming2013}
\begin{equation}
\begin{aligned}
C=E\left\{\log _{2}\left[\operatorname{det}\left(\mathbf{I}+\frac{\gamma}{N_{t}} \mathbf{R} \mathbf{H}_{w} \mathbf{H}_{w}^{H}\right)\right]\right\},
\end{aligned}
\end{equation}%$\mathbf{H}_{ N_{r} \times N_{t}}$
where $E$ represents the mathematical expectation, $^H$ is the Hermitian operator, the covariance matrix $\mathbf{R}$ equals $\boldsymbol{\Phi}$ in the Clarke model (ideal antenna efficiencies), and $\mathbf{I}$ is the identity matrix. Moreover, $N_{t}$ and $N_{r}$ are the number of transmitting and receiving antennas {($N_{t}=N_{r}$ is assumed here for the convenience of analysis)}, $\gamma$ is the fixed total signal-to-noise ratio (SNR), and the entries of $\mathbf{H}_{w}$ are independent and identically distributed (i.i.d.) complex Gaussian variables. {Usually, the $\mathbf{H}_w$ is normalized by making $\|{\mathbf{H}_w}\|_F^{2}=N_{t} N_{r}$ for fair comparisons \cite{tse2005fundamentals, Loyka2009}, where the $N_{t}N_{r}$ represents the array gain (or power gain) of the MIMO channel. {This is a frequently used assumption in MIMO communications, the antennas are simplified as isotropic point sources, and the element spacing is an integer multiple of $\lambda_0/2$. Under this assumption, the array gain is simply equal to the number of antennas (assuming the antennas are lossless) .} However, the receiving array gain of an aperture-constrained array will not further increase when the element spacing becomes smaller than $\lambda_0/2$.} Hence, the $\mathbf{H}_w$ is  normalized by making $E\left\{\|{\mathbf{H}_w}\|_F^{2}\right\}=N_{t} N_{r}$ when the element spacing is larger than half wavelength, and making  $E\left\{\|{\mathbf{H}_w}\|_F^{2}\right\}=N_{t}N_{\lambda_0/2}$ otherwise \cite{ShuaiOJAP, Loyka2009}, where $N_{\lambda_0/2}$ is the number of antennas at $\lambda_0/2$ element spacing.
\subsection{Kronecker model}
For practical antenna arrays, the deformation of radiation patterns and the decrease of radiation efficiencies caused by mutual coupling are required to be taken into consideration \cite{Kildal2004, xiaoming2013}. The covariance matrix $\mathbf{R}$ is constructed by the entry-wise product between the correlation matrix $\boldsymbol{\Phi}$ and the embedded efficiency matrix $\boldsymbol{\Xi}$, i.e.,
\begin{equation}
\mathbf{R}=  \boldsymbol{\Phi}\circ\boldsymbol{\Xi}.
\end{equation}
The correlation matrix incorporating mutual coupling is \cite{xiaoming2017}
\begin{equation}
\boldsymbol{\Phi}=\left[\begin{array}{cccc}
1 & \rho_{12} & \cdots & \rho_{1 N_{r}} \\
\rho_{12}^{*} & 1 & \cdots & \rho_{2 N_{r}} \\
\vdots & \vdots & \ddots & \vdots \\
\rho_{1 N_{r}}^{*} & \rho_{2 N_{r}}^{*} & \cdots & 1
\end{array}\right],
\end{equation}
where
\begin{equation}
\rho_{mn}=\frac{\oint G_{mn}(\Omega) \mathrm{d} \Omega}{\sqrt{\oint G_{mm}(\Omega) \mathrm{d} \Omega} \sqrt{\oint G_{nn}(\Omega) \mathrm{d} \Omega}},
\end{equation}
with
\begin{equation}
G_{m n}(\Omega)=\kappa E_{\theta m}(\Omega) E_{\theta n}^{*}(\Omega) P_{\theta}(\Omega)+E_{\phi m}(\Omega) E_{\phi n}^{*}(\Omega) P_{\phi}(\Omega),
\end{equation}
$ E_{\theta}(\Omega)$ and $ E_{\phi}(\Omega)$ are the $\theta-$ and $\phi-$polarized embedded radiation patterns, ${P}(\Omega)$ is the angular power spectrum, and $\kappa$ is the cross-polarization discrimination (XPD). The XPD is taken as 1 (polarization-balanced), and  ${P}(\Omega)$ could be used for characterizing different angular spreads. The embedded efficiency matrix $\boldsymbol{\Xi}$ is
\begin{equation}
\boldsymbol{\Xi}=\sqrt{\mathbf{e}}\sqrt{\mathbf{e}}^{T},
\end{equation}
with
\begin{figure}[ht!]
	\centering
	\includegraphics[width=3.4in]{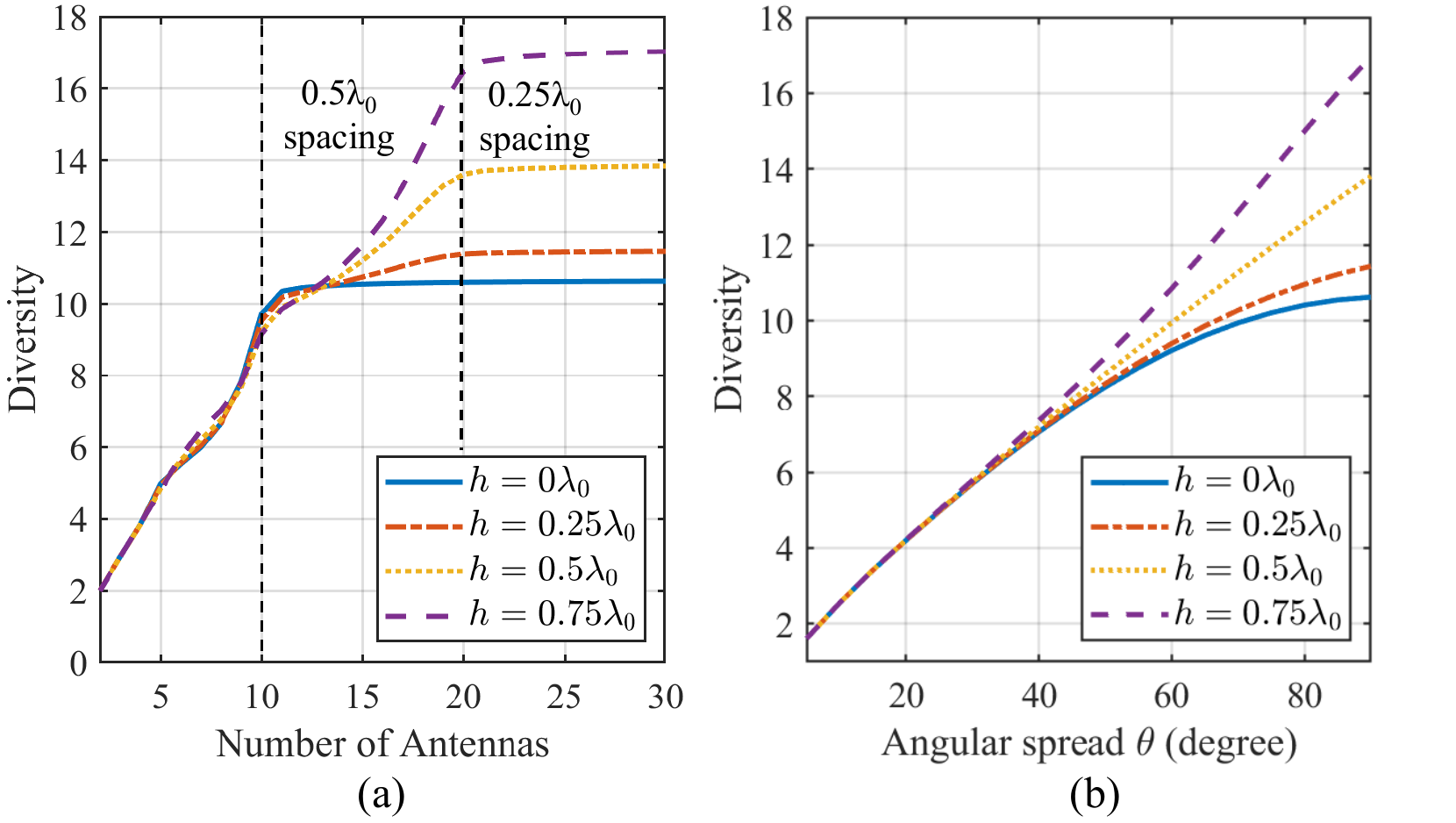}
	\caption{Diversity measure based on the 3-D Clarke model, and the array length is fixed as 5$\lambda_0$. (a) Diversities of the 3-D arrays with different $h$ and antenna numbers. (b) Diversities of the 3-D arrays with different $h$ and angular spreads, where the antenna number is fixed as 25.}%=d_x\times N=120
	\label{4}
\end{figure}%=======figure==========%
\begin{figure}[ht!]
	\centering
	\includegraphics[width=3.4in]{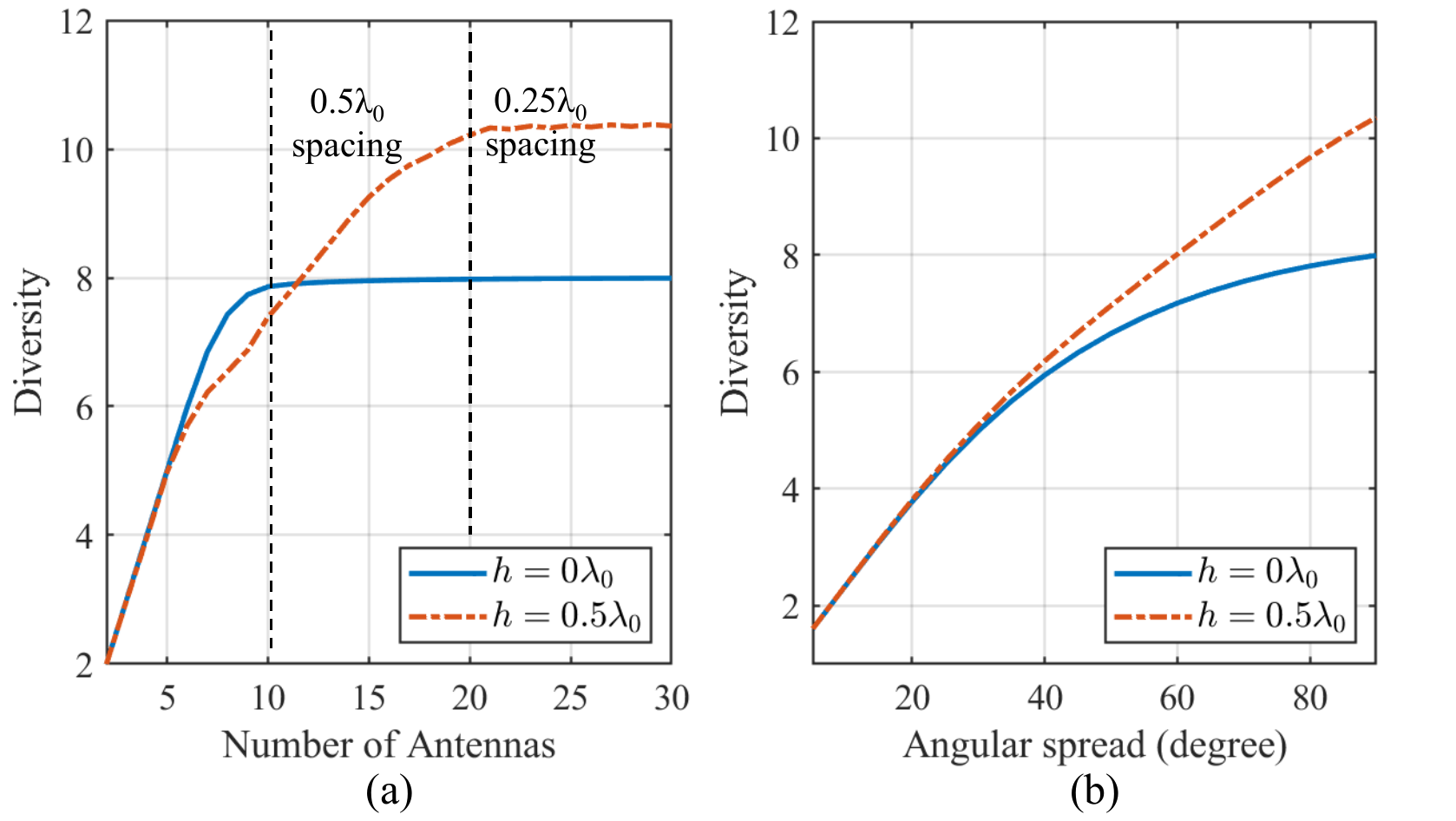}
	\caption{Diversity measure based on the Kronecker model, and the array length is fixed as 5$\lambda_0$. (a) Diversities of the 3-D arrays when $h=0$ and $h=0.5$$\lambda_0$. (b) Diversities of the 3-D arrays with different $h$ and angular spreads, where the antenna number is fixed as 25.}%=d_x\times N=120
	\label{5}
\end{figure}%=======figure==========%
\begin{equation}
\mathbf{e}=\left[e^{e m b }_1, e^{e m b }_2,  \cdots, e^{e m b }_{N_{r}}\right]^{T},
\end{equation}
where the embedded radiation efficiency of the $n$th antenna is calculated by the $S$ parameters assuming negligible ohmic loss \cite{Kildal2016}
\begin{equation}
e^{e m b}_{n}=1-\left|S_{1n}\right|^{2}-\left|S_{2n}\right|^{2}-\cdots -\left|S_{N_rn}\right|^{2}.
\end{equation}
{In the characterization of practical arrays, the power coupled to other ports is not regarded as the radiated power \cite{kildal2015foundations}.}
\section{Theoretical analysis}
We start with the fundamental Clarke model, wherein antennas are approximated as isotropic point sources, and mutual couplings are disregarded. In this section, the lengths of all the arrays are fixed as 5$\lambda_0$, and a height difference $h$ is introduced between the nearby antennas for realizing 3-D array (see Fig. \ref{1}), and the 3-D array will degenerate into a 2-D array when $h=0$. The diversities of 3-D arrays in different scenarios are investigated and presented in Fig. \ref{4}. In Fig. \ref{4}(a), it can be observed that the diversity in 2-D case ($h=0$) will not further increase when element spacing reaches nearly 0.5$\lambda_0$, which is the fundamental DOF limit of conventional holographic MIMO communications. However,  the limit can be surpassed when a height difference between the nearby antennas is introduced, and the diversity keeps increasing until nearly 0.25$\lambda_0$ element spacing. Moreover, the diversities under different angular spreads are presented in Fig. \ref{4}(b), where the increment of diversity brought by the 3-D configuration is positively related to the angular spread. 
\begin{figure}[ht!]
	\centering
	\includegraphics[width=3in]{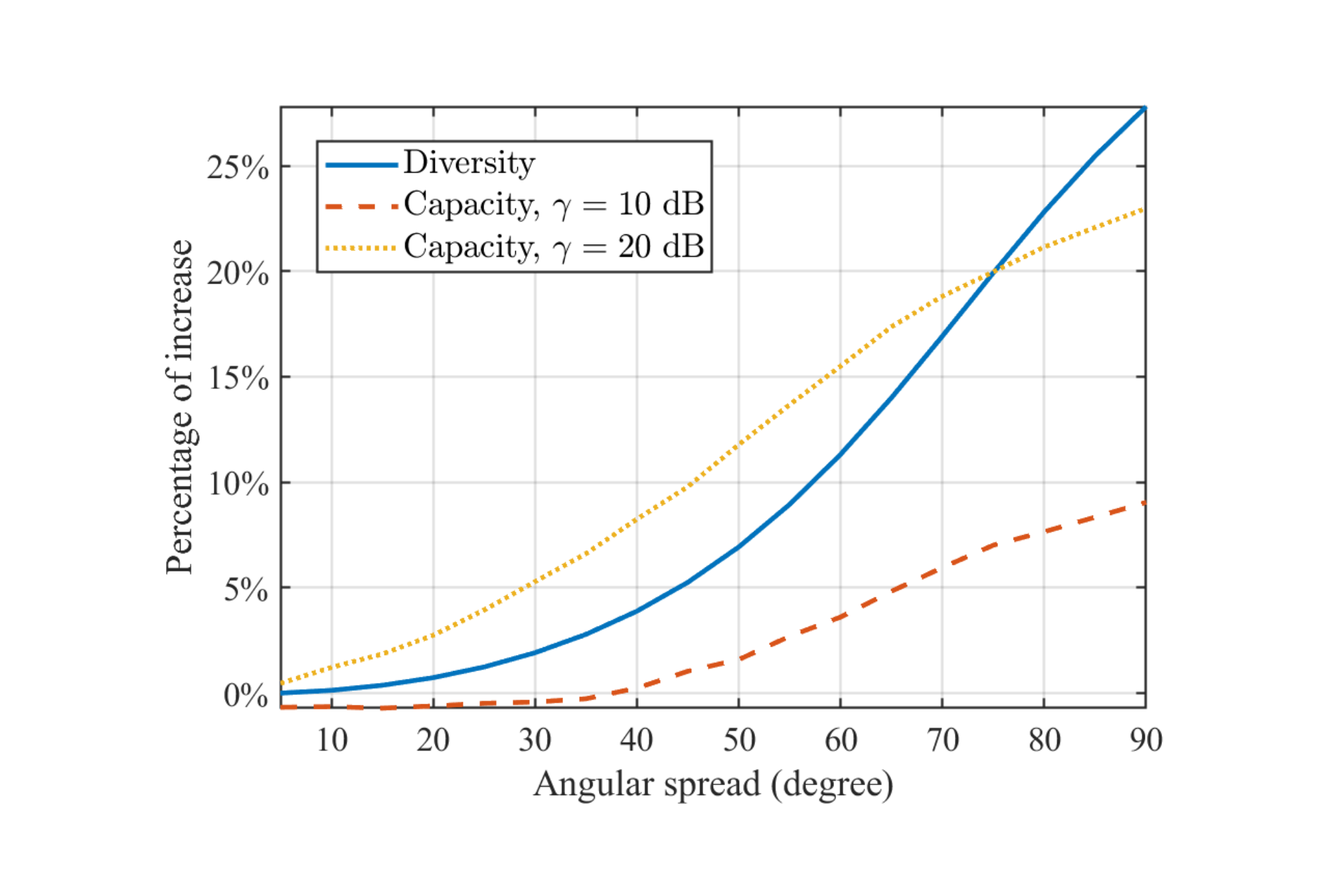}
	\caption{Compared to using the 2-D array with the same aperture size, the percentages of increases in diversity and capacity by using the 3-D array.}%=d_x\times N=120
	\label{6}
\end{figure}%=======figure==========%

The Kronecker model can be used for taking into account the distorted antenna radiation patterns. Rather than using the ideal isotropic radiation patterns, we first take into account the distortion of radiation patterns due to the 3-D antenna array topology, where the radiation patterns of the lower and upper antennas would be different. With a large height difference, side lobes are almost unavoidable for upper antennas according to the antenna array theory \cite{balanis2015antenna}. A 0.5$\lambda_0$ height difference is taken here for balancing the antenna performance and DOF benefits, which is acceptable for a practical MIMO array. The radiation patterns of the upper and lower antennas are exported for performing the theoretical analyses here. The corresponding antenna designs and radiation patterns can be found in Fig. \ref{8}(b-c) and Fig. \ref{11}. 

Based on the Kronecker model, the diversities of 3-D arrays are calculated and presented in Fig. \ref{5}, showing that the DOF limit can be broken by using the 3-D configuration, and the increment brought by the 3-D array is largely dependent on the angular spread. {It should be noticed that the diversities of 3-D arrays could be slightly lower than that of 2-D arrays without enough antennas, as shown in Fig. \ref{5}(a). In this situation, the performance of the 2-D array is close to saturation, while the space occupied by the 3-D array is still not well sampled. {The performance limit should always be reached or exceeded with a sufficient antenna number (i.e., holographic MIMO), and there may be few benefits from the 3-D array when the antenna number is not enough.} For a better illustration, we plot the percentage of increases in diversity and capacity brought by the 3-D array compared to the 2-D array with the same aperture size in Fig. \ref{6}. 
	\begin{figure}[ht!]
		\centering
		\includegraphics[width=3in]{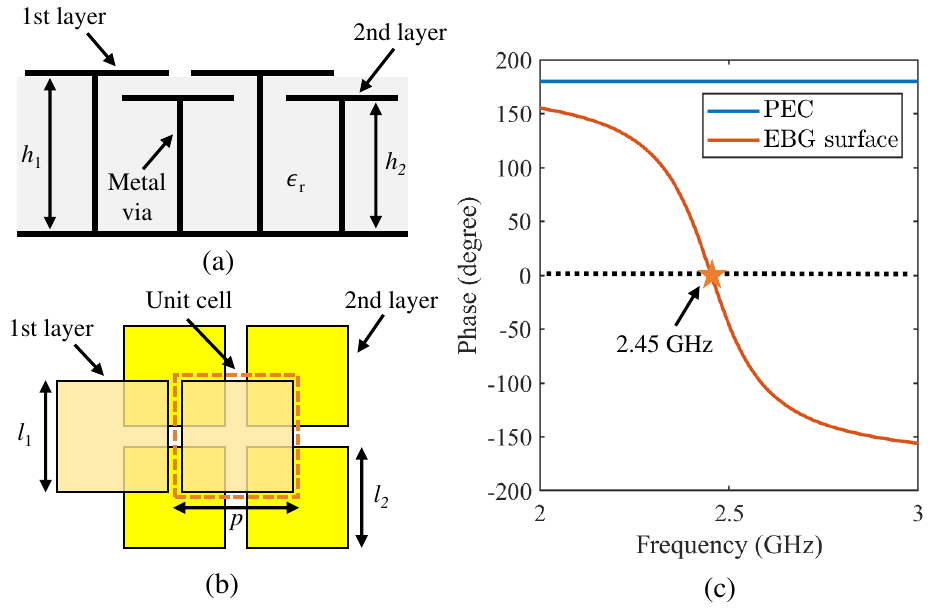}
		\caption{Unit cell of the EBG surface for enabling the proposed 3-D array topology. (a) Side view of the double-layer EBG units, $h_1= 1.83$ mm, $h_2= 1.57$ mm. The relative dielectric constant of the substrate is $\epsilon_r=2.2$. (b) Top view of the double-layer EBG units, $l_1=10.5$ mm, $l_2=9.5$ mm, $p=11$ mm. (c) Reflecting phase of the EBG surface obtained from full-wave simulations, and a $0^{\circ}$ reflecting phase can be realized at 2.45 GHz.}%=d_x\times N=120
		\label{7}
	\end{figure}%=======figure==========%
	\begin{figure}[ht!]
		\centering
		\includegraphics[width=3.4in]{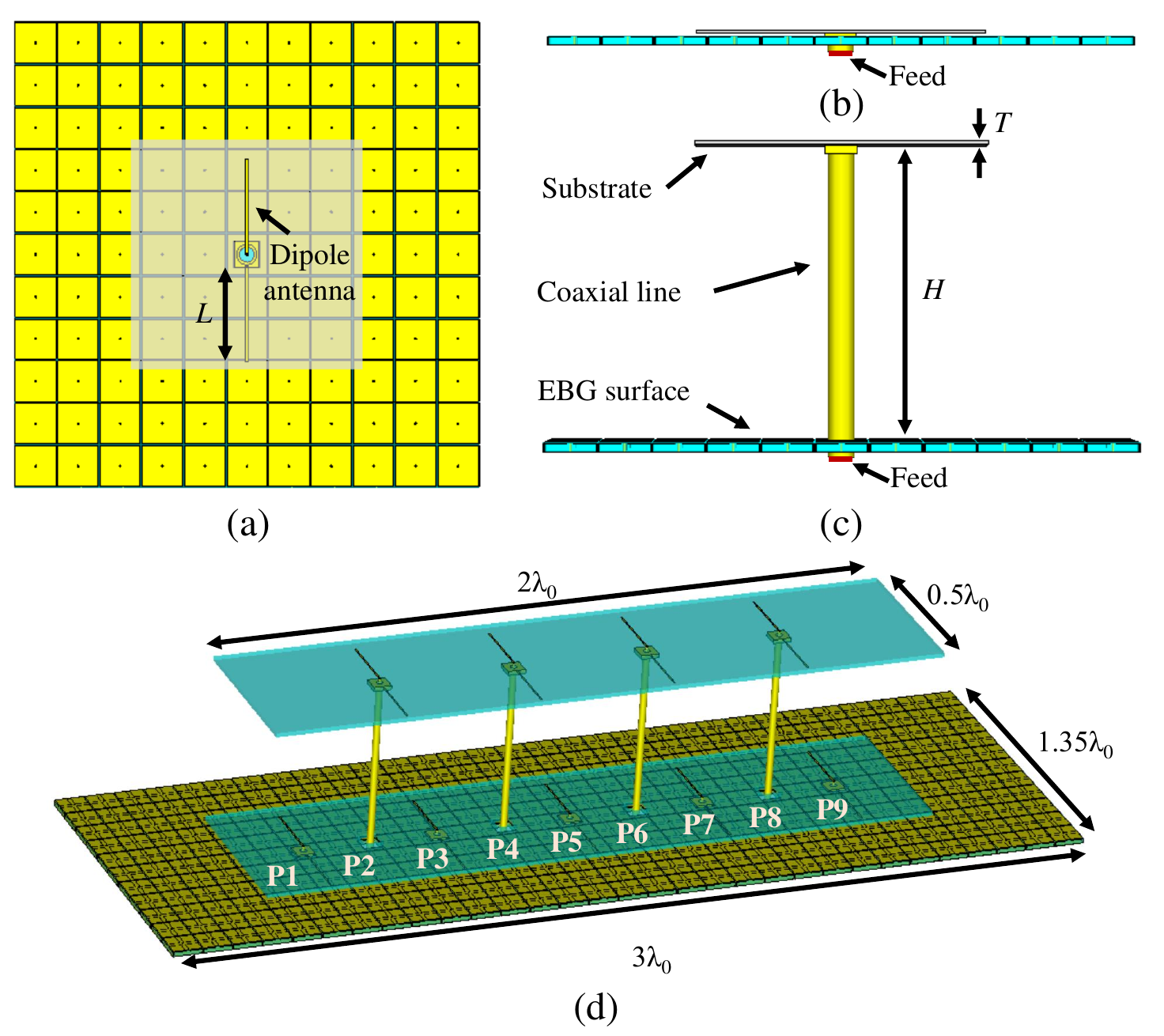}
		\caption{Configuration of the 3-D antenna array. (a) Top view of a single antenna element, $L=48$ mm. (b) Side view of the lower antenna element. (c) Side view of the upper antenna element, $H=61$ mm, $T=1.57$ mm. (d) Perspective view of the 3-D antenna array.}%=d_x\times N=120
		\label{8}
	\end{figure}%=======figure==========%
	\begin{figure}[ht!]
		\centering
		\includegraphics[width=3.4in]{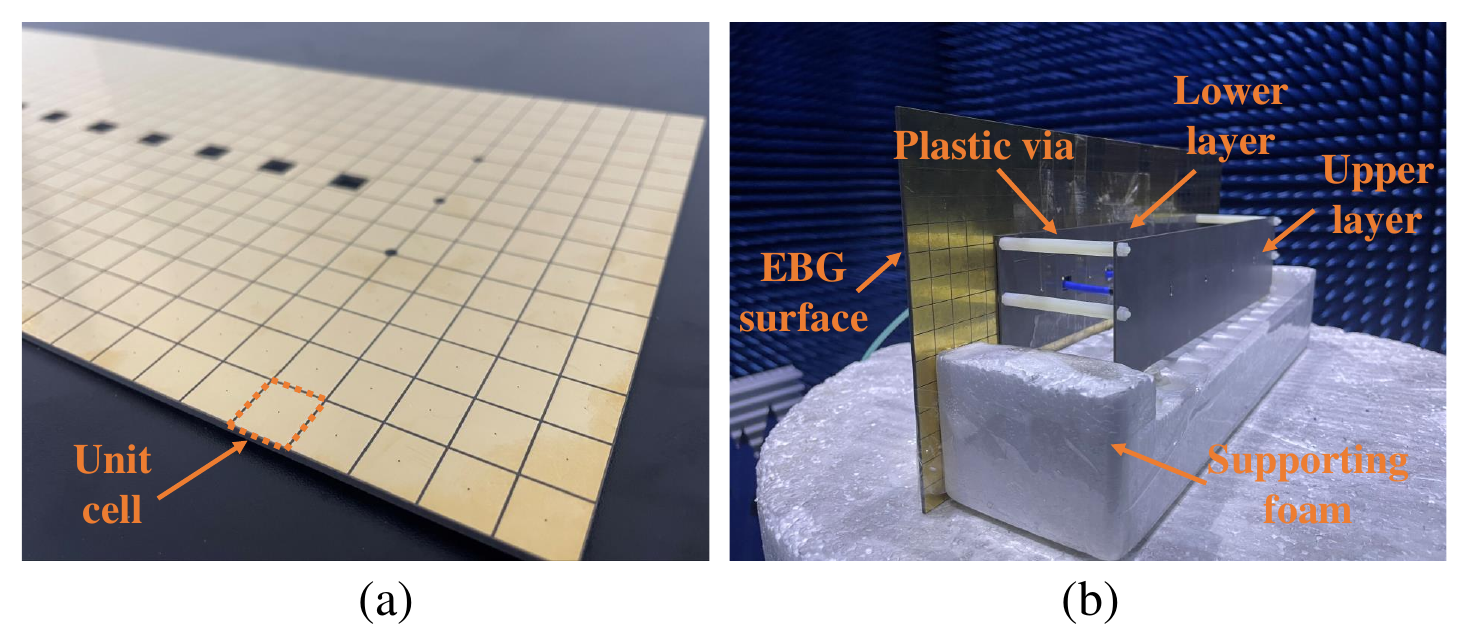}
		\caption{Fabricated samples and experimental setup. (a) Fabricated EBG surface. (b) Fabricated 3-D antenna array and testing environment.}%=d_x\times N=120
		\label{9}
	\end{figure}%=======figure==========%
	\begin{figure}[ht!]
		\centering
		\includegraphics[width=3in]{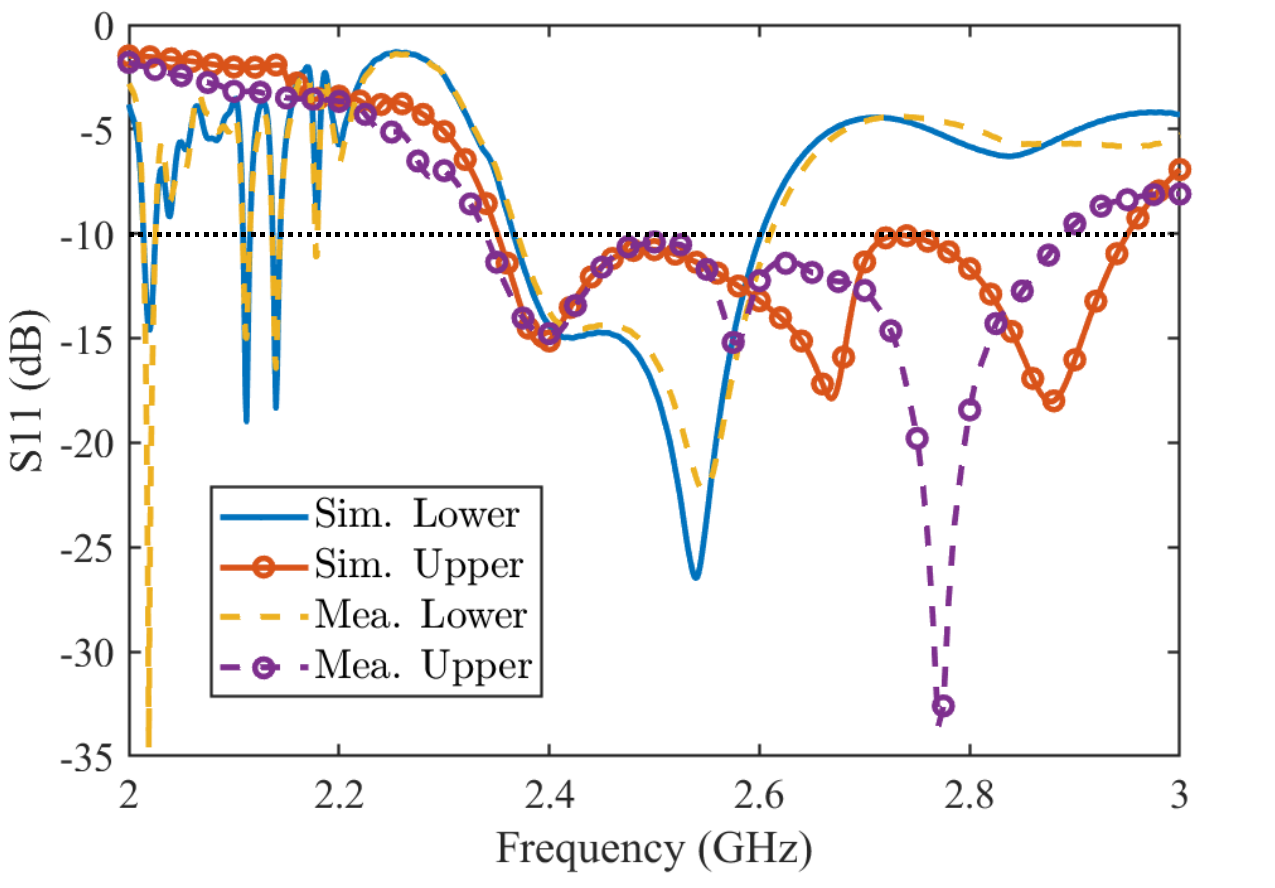}
		\caption{Simulated and measured reflection coefficients of the isolated lower and upper antennas. }%=d_x\times N=120
		\label{10}
	\end{figure}%=======figure==========%
	\begin{figure}[ht!]
		\centering
		\includegraphics[width=3in]{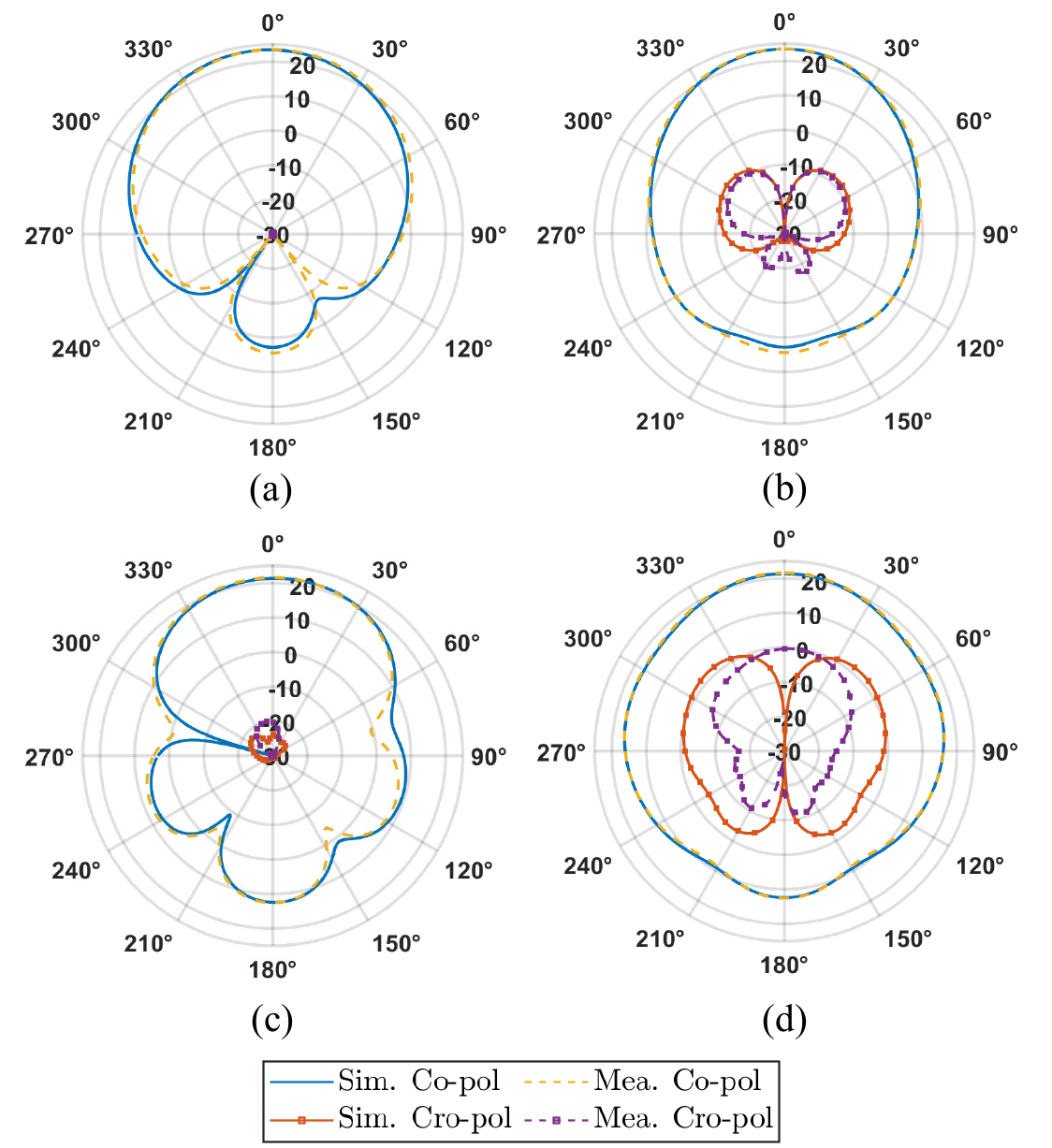}
		\caption{Simulated and measured electric-field patterns of the isolated lower and upper antennas at far field (reference distance is taken as 1 m), and the unit is dB(V/m). {The ‘Co-pol’ and ‘Cro-pol’ represent the co-polarized and cross-polarized patterns, respectively.} (a-b) E-plane and H-plane patterns of the lower antenna. (c-d) E-plane and H-plane patterns of the upper antenna.}%=d_x\times N=120
		\label{11}
	\end{figure}%=======figure==========%	
	Theoretically, i.e., ideal decouplings are made, the diversity can be increased by $27\%$, and the capacity can be increased by $9\%$ ($\gamma=10$ dB) and $22.5\%$ ($\gamma=20$ dB) in an isotropic multi-path environment (angular spread is $90^{\circ})$. In practical applications, the angular spread may become smaller. When the angular spread is $60^{\circ}$, the diversity can still be increased by $12\%$, and the capacity can be increased by $4\%$ ($\gamma=10$ dB) and $15.5\%$ ($\gamma=20$ dB). Hence, the theoretical analyses put forth in this study strongly suggest that the proposed 3-D configuration successfully transcends the DOF limit, showcasing its potential application for enhancing capacity across a multitude of scenarios.
	\section{Numerical and experimental verifications}
	\subsection{Design of the 3-D  array}
	\begin{figure*}[ht!]
		\centering
		\includegraphics[width=6in]{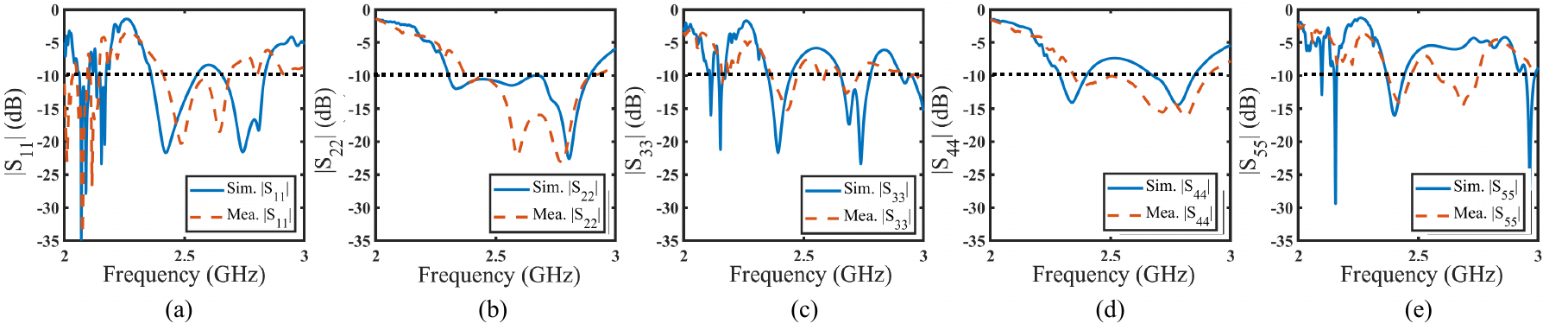}
		\caption{Simulated and measured reflection coefficients of the antennas in a 3-D array. (a-e) Reflection coefficients of the antennas P1 - P5.}%=d_x\times N=120
		\label{12}
	\end{figure*}%=======figure==========%
	\begin{figure*}[ht!]
		\centering
		\includegraphics[width=6in]{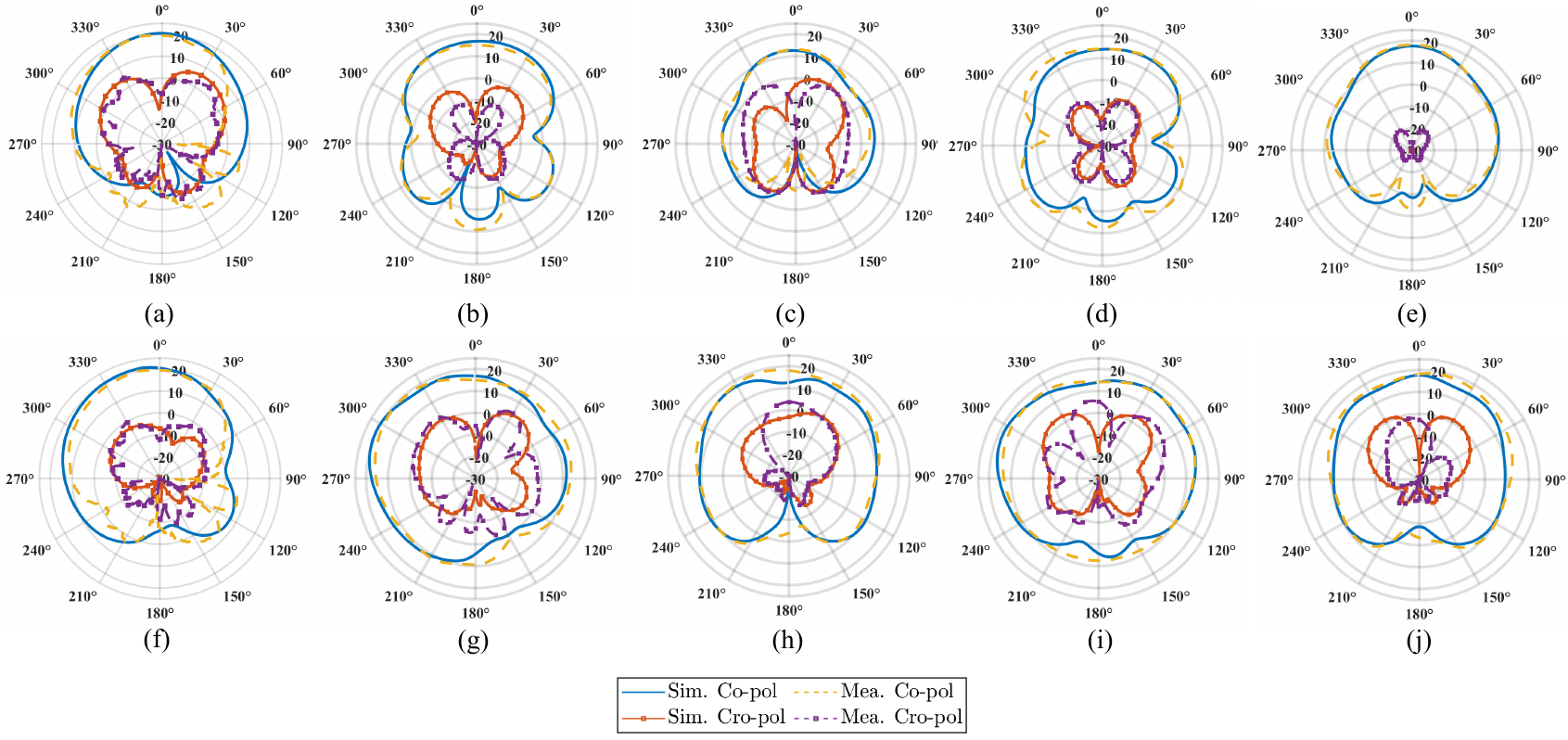}
		\caption{Simulated and measured far-field patterns of the antennas in a 3-D array, the unit is dB(V/m). (a-e) E-plane patterns of the antennas P1 - P5. (f-j) H-plane patterns of the antennas P1 - P5.}%=d_x\times N=120
		\label{13}
	\end{figure*}%=======figure==========%
	As a proof of concept, a practical 3-D array enabled by an EBG surface is designed for verifying the above theoretical analyses. To create the 0.5$\lambda_0$ height difference while maintaining acceptable antenna performances, a $0^{\circ}$ reflecting phase should be introduced by using the EBG surface at the working frequency. The design of EBG surfaces has been well discussed in literature \cite{yang2009electromagnetic}, and a double-layer EBG surface is used here for realizing the desired functions \cite{abedin2005effects}. The diagram of the utilized EBG structure is plotted in Fig. \ref{7}(a-b), and a $0^{\circ}$ reflecting phase can be realized at 2.45 GHz in Fig. \ref{7}(c). To reduce the complexities of design, simulation, and fabrication, a microstrip dipole antenna is used here for facilitating the procedure of verification, as shown in Fig. \ref{8}(a). The full-wave models of the isolated upper and lower antennas are presented in Fig. \ref{8}(a-c), and a perspective view of the constructed 3-D array can be found in Fig. \ref{8}(d). Furthermore, the proposed 3-D array in Fig. \ref{8}(d) is fabricated and fully measured, the fabricated samples and experimental setup are demonstrated in Fig. \ref{9}.
	\subsection{Antenna performances of the 3-D array}
	\subsubsection{Isolated antenna}
	The reflection coefficients and far-field radiation patterns of the isolated upper and lower antennas are first investigated, as shown in Fig. \ref{10}. {The simulated and measured results agree well with each other, and the reflection coefficients are below -10 dB near the designed working frequency, which fulfills the requirements for array elements.} Moreover, the electric-field patterns of the isolated antennas at far field (reference distance is taken as 1 m) are demonstrated in Fig. \ref{11}. It can be observed that the lower antenna maintains a regular radiation pattern \cite{yang2003reflection}, while the upper antenna has side lobes due to the introduced 0.5$\lambda_0$ height difference. {The side lobes in the upper antenna are almost inevitable, since its radiation pattern can be regarded as the end-fire radiation of a two-element in-phase array with 1$\lambda_0$ distance. To create a larger height difference, directly increasing the height difference is detrimental to radiation patterns; however, a multilayer 3-D array would be more feasible.} Moreover, the performances of antennas in the 3-D array can be further improved by the antenna and decoupling designs.
	\subsubsection{Antennas in the 3-D array}
	The $S$ parameters and embedded radiation patterns of the antennas in a practical 3-D array are simulated and measured. Specifically, the embedded radiation pattern of one antenna in the array is obtained by exciting this antenna while making all the other antennas well-matched. The results of the first five antennas, i.e., P1-P5 in Fig. \ref{8}(d), are given considering the symmetric structure of 3-D array. {The simulated and measured reflection coefficients are plotted and compared in Fig. \ref{12}. Near the desired frequency, the reflection coefficients are mostly below -10 dB, some deviations can be attributed to the fabrication and measurement errors. In Fig. \ref{13}, the simulated and measured embedded radiation patterns of the antennas in the 3-D array are given, where some parts of radiation patterns are deformed due to the mutual coupling at small element spacings.}
	
	\begin{figure}[ht!]
		\centering
		\includegraphics[width=2.8in]{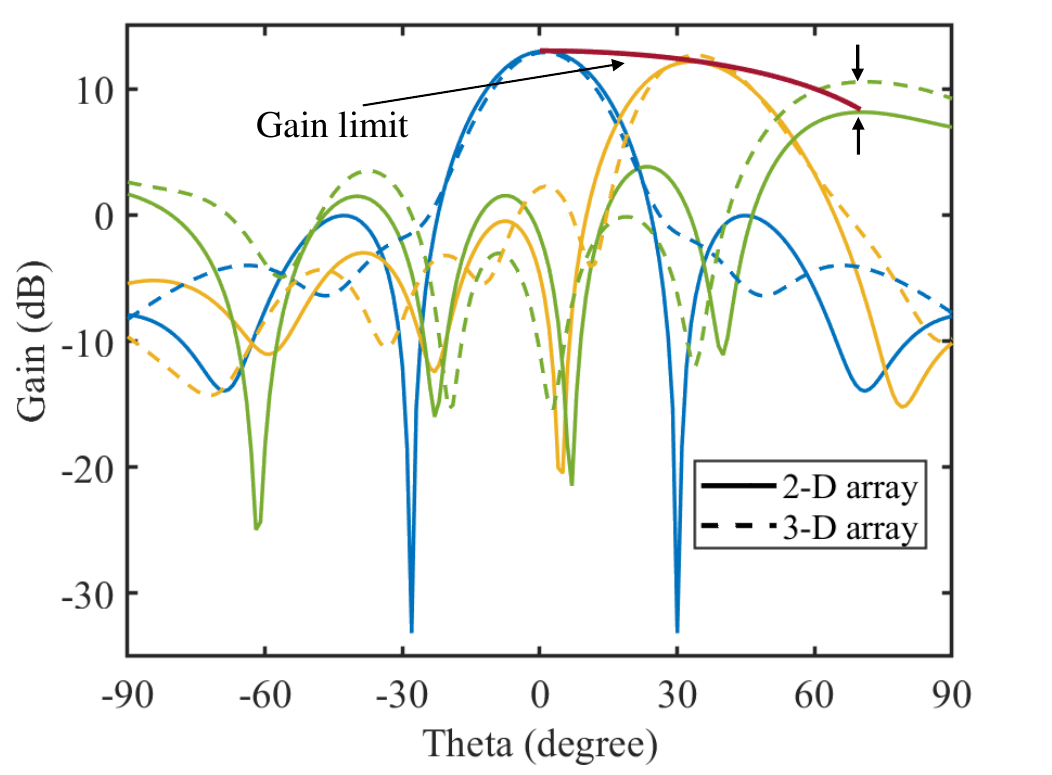}
		\caption{{Beamforming performances of the 2-D and 3-D arrays. The blue, orange, and green lines represent the gain patterns at the target beamforming angles of 0, 35, and 70 degrees, respectively. The solid and dotted lines represent the gain patterns of the 2-D and 3-D arrays. The red line denotes the theoretical gain limit of the 2-D array, which can be exceeded by the 3-D array at large scanning angles.}}%=d_x\times N=120
		\label{14}
	\end{figure}%=======figure==========%
	\begin{figure}[ht!]
		\centering
		\includegraphics[width=3in]{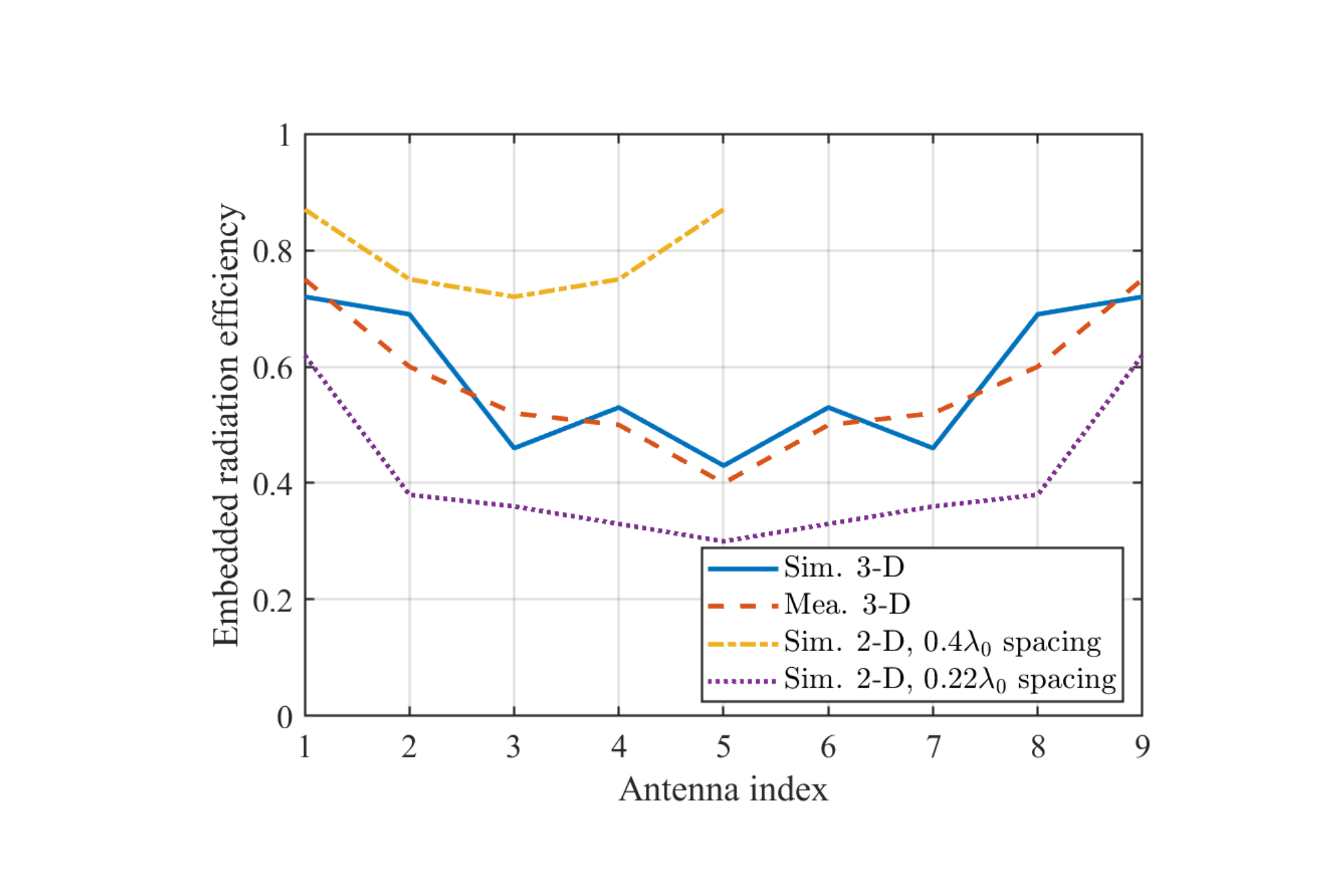}
		\caption{Embedded radiation efficiencies of the antennas in different types of arrays.}%=d_x\times N=120
		\label{15}
	\end{figure}%=======figure==========%
	{From the perspective of beamforming, the 3-D array will have better gain than the 2-D array especially at large scanning angles, because the corresponding projection area is larger. The gain limit of a 2-D array is given by \cite{balanis2015antenna}}
	\begin{equation}G=\frac{4 \pi A}{\lambda^2} \cos \theta_0,\end{equation}
	where $A$ is the area of the array, and $\cos\theta_0$ characterizes the decrease of projection area due to the increase of $\theta_0$. The 3-D array is expected to have a larger projection area, especially at large scanning angles due to the vertical height difference. Therefore, a well-designed 3-D array could break the gain limit of the 2-D array at large $\theta_0$ angles, as shown in Fig. \ref{14}.
	
	The embedded radiation efficiencies of the antennas are calculated from $S$ parameters according to (10), which are drawn and compared to the efficiencies in the 2-D arrays at different element spacings (i.e., different antenna numbers) in Fig. \ref{15}. It can be found that the efficiencies in the 3-D array are generally lower than that in the 2-D array at 0.4$\lambda_0$ spacing, but higher than that in the 2-D array at 0.22$\lambda_0$ spacing (with the same antenna number as the 3-D array). The latter indicates that Hannan's efficiency limit for a planar array \cite{Hannan1964} is in fact broken by introducing the height difference in the 3-D array. 
	\begin{figure}[ht!]
		\centering
		\includegraphics[width=3.38in]{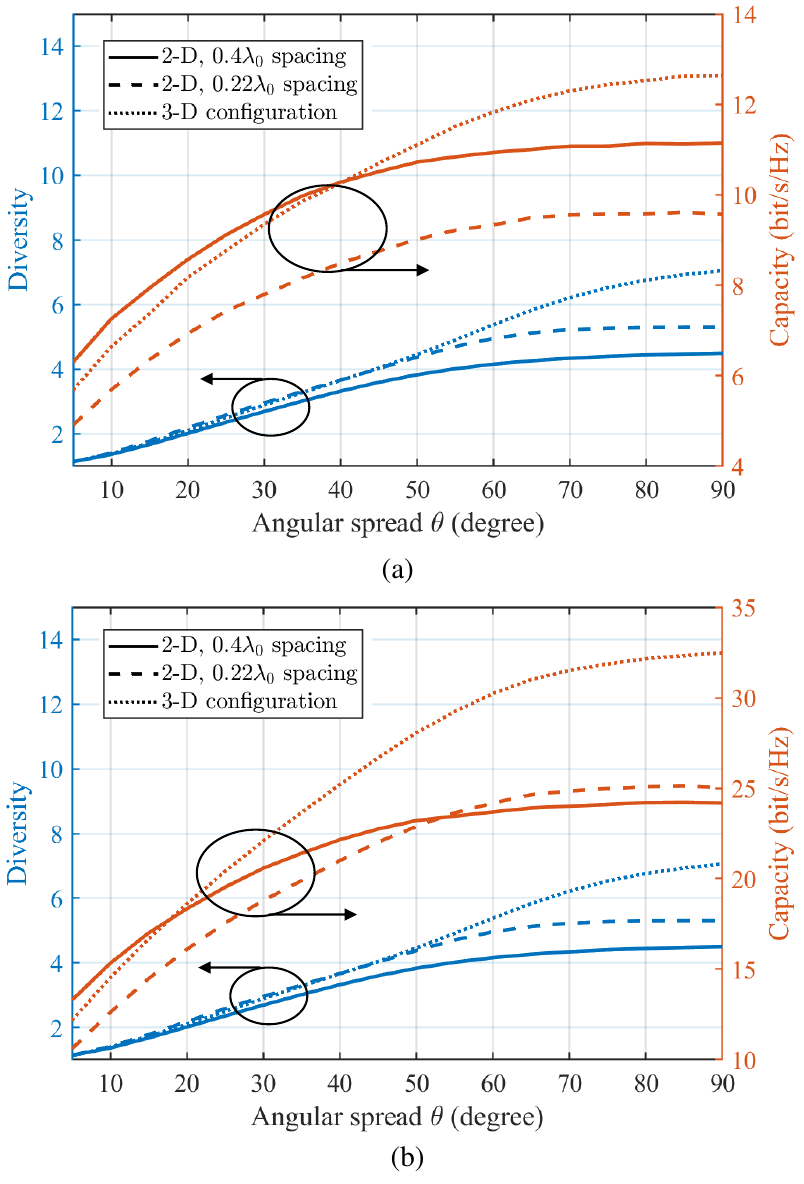}
		\caption{Diversity and capacity benefits of the 3-D array compared to the 2-D array with the same aperture size, and the array length along the $x$-axis is 2$\lambda_0$. (a) SNR $\gamma=10$ dB. (b) SNR $\gamma=20$ dB.}%=d_x\times N=120
		\label{16}
	\end{figure}%=======figure==========%
	\begin{figure}[ht!]
		\centering
		\includegraphics[width=3.4in]{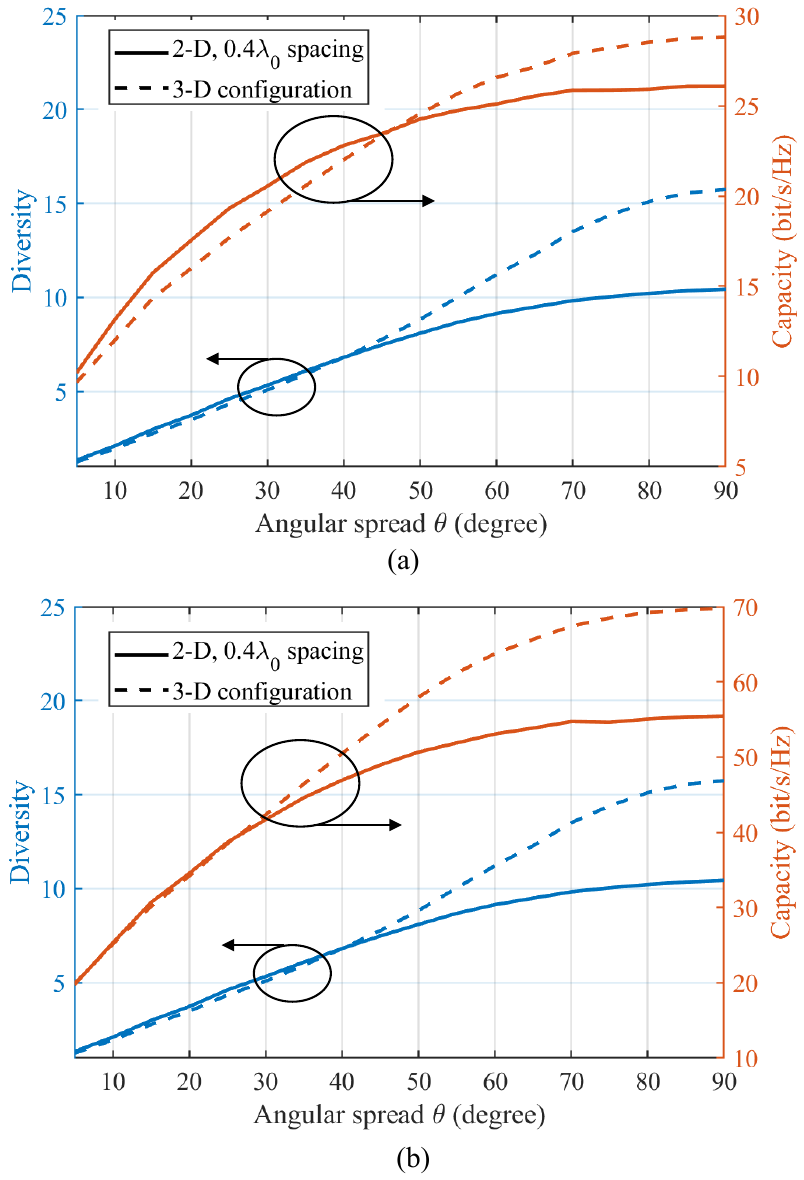}
		\caption{Diversity and capacity benefits of the 3-D array compared to the 2-D array with the same aperture size, and the array length along the $x$-axis is 5$\lambda_0$. (a) SNR $\gamma=10$ dB. (b) SNR $\gamma=20$ dB.}%
		\label{17}
	\end{figure}%=======figure==========%
	\subsection{MIMO performances of the 3-D array}
	\subsubsection{2$\lambda_0$ array in Rayleigh channel}
	With the above results, i.e., embedded radiation patterns and efficiencies, the diversity and capacity of the 3-D array can be calculated and compared to the 2-D arrays with the same aperture size by using (3-10).  Three types of arrays with 2$\lambda_0$ array length are first considered, including the proposed 3-D array, the 2-D array at 0.4$\lambda_0$ element spacing (i.e., only the lower layer of the 3-D array), and the 2-D array at 0.22$\lambda_0$ element spacing (with the same antenna number as the 3-D array). The results at different SNR levels ($\gamma =$ 10 and 20 dB) are depicted in Fig. \ref{16}(a) and Fig. \ref{16}(b), {and the arrows in the figures indicate the lines corresponding to the diversity and capacity.} It can be observed that the increments of diversities and capacities are close to the theoretical analyses in Section III.  To be specific, compared to the 2-D array, the diversity is increased by $33\%$, and the capacity is increased by $13\%$ ($\gamma =$ 10 dB) and $30\%$ ($\gamma =$ 20 dB) in isotropic multi-path environment. When the angular spread is $60^\circ$, the diversity is increased by $18\%$, and the capacity is increased by $9\%$ ($\gamma =$ 10 dB) and $19\%$ ($\gamma =$ 20 dB). Particularly, for a regular 2-D array, the diversity may only slightly increase due to the mutual coupling after reaching the DOF limit, but the efficiency will dramatically decrease following Hannan's limit \cite{Hannan1964}. Therefore, placing too many antennas into a constrained planar area will generally be harmful to capacity, as shown by the 0.22$\lambda_0$ and 0.4$\lambda_0$ spacing 2-D arrays in Fig. \ref{16}(a). 
	\subsubsection{5$\lambda_0$ array in Rayleigh channel}
	We also investigate the performance of a larger 3-D array in Fig. \ref{17}, with the length of 5$\lambda_0$ along the $x-$axis. In a large-scale array, the radiation patterns of the antennas in the middle region are similar \cite{kildal2015foundations}, and the influence of edge antennas would be little. Compared to the 2-D array at 0.4$\lambda_0$ spacing, the capacity is increased by $12\%$ ($\gamma =$ 10 dB) and $26\%$ ($\gamma =$ 20 dB) in isotropic multi-path environment, and increased by $8\%$ ($\gamma =$ 10 dB) and $20\%$ ($\gamma =$ 20 dB) when the angular spread is $60^\circ$. {In order to analyze the effects of mutual coupling, the percent of increase is drawn in Fig. \ref{18} similar to the theoretical analysis in Fig. \ref{6}. It can be observed that the increments of capacities are slightly larger than the theoretical analyses in Section III at large angular spreads, because the large-angle radiations are stronger (indicating more DOF benefits are brought by the 3-D array configuration). However, the mutual coupling will degrade the performance of the 3-D array at low SNR and small angular spreads (the capacity enhancement brought by the extra DOF is not significant), also bring troubles for beamforming. 
		\begin{figure}[ht!]
			\centering
			\includegraphics[width=3in]{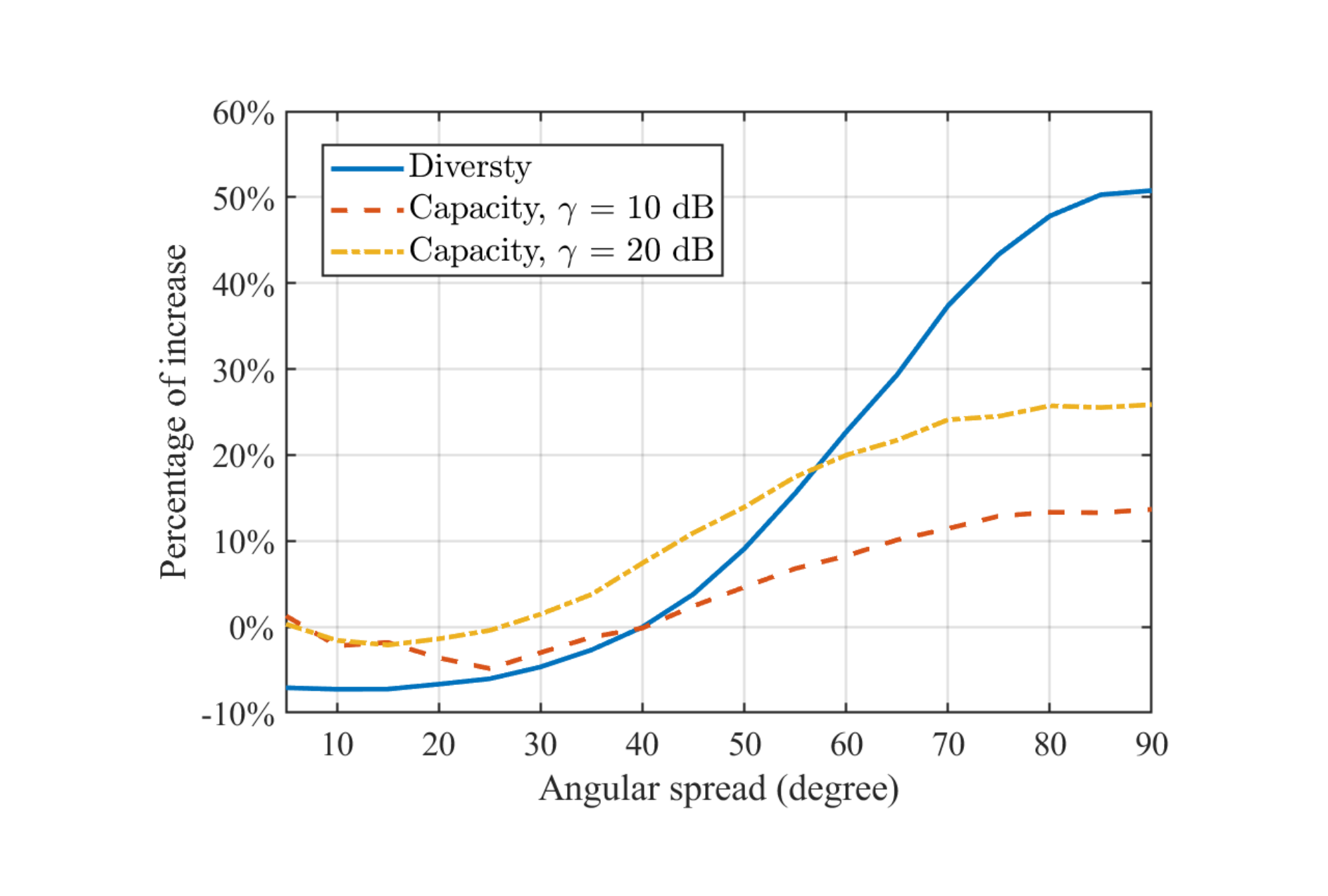}
			\caption{{Compared to using the 2-D array with the same aperture size, the percentages of increases in diversity and capacity by using the 3-D array configuration in a practical case.}}%=d_x\times N=120
			\label{18}
		\end{figure}%=======figure==========%
		\begin{figure}[ht!]
			\centering
			\includegraphics[width=3.4in]{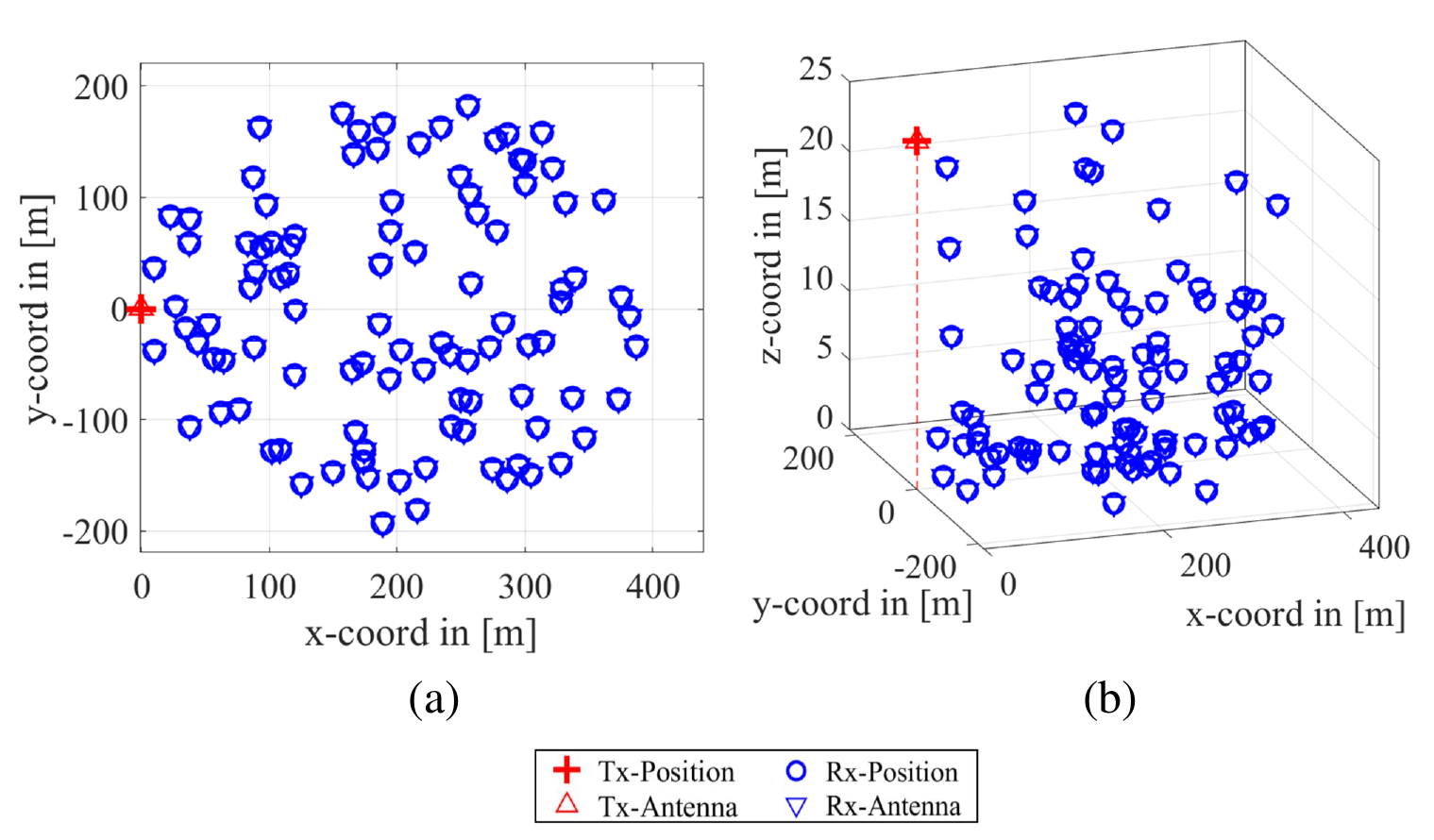}
			\caption{Configurations of 2-D and 3-D 3GPP channels with 100 users. (a) 2-D 3GPP channel. (b) 3-D 3GPP channel.}%
			\label{19}
		\end{figure}	
		{It also should be noticed that the 3-D array provides marginal benefits over the 2-D array in far-field LOS scenarios \cite{hu2018beyond, Song2015}, which is in consistent with our analyses at small angular spreads.} Therefore, the benefit of 3-D array is highly dependent on the SNR level, specific multi-path environments, and antenna designs (radiation patterns), while the obtained results reveal the superiorities of the 3-D array over the 2-D array with the same aperture size in many scenarios.}
	\subsubsection{3GPP channel}
	To further investigate the MIMO performances of the 3-D array in practical communication environments rather than in the ideal Rayleigh channel, the open-source software QuaDRiGa is used for comparing the performances of 2-D and 3-D arrays in 3GPP channels \cite{Quadriga}. Particularly, two single-cell scenarios are considered, including the \textit{2-D 3GPP urban macro (UMa)} and \textit{3-D 3GPP UMa} scenarios \cite{Quadriga}. The configurations of the 3GPP models are demonstrated in Fig. \ref{19}, where the users (marked by blue circles) are distributed in a circular area with a radius of 200 m, and the base station (marked by a red cross) is placed on the left side of the circular area with the height of 25 m. 
	
	The users are arranged in accordance with the 3GPP assumptions\cite{3GPP}, where 80$\%$ of them are situated indoors at different floor levels for the 3-D scenario and all set outdoors (1.5 m height) for the 2-D scenario. The embedded radiation patterns of the proposed 2$\lambda_0$ 3-D array are imported as a transmitting array, while the users are set as omnidirectional antennas. {After that, the channels of the paths created by different clusters are generated and summed for obtaining the wanted correlation matrix. Following (2-10), the diversity and capacity can be obtained for investigating the performances of 3-D arrays in geometry-based spatial correlation channels.} The diversities and capacities ($\gamma$ = 20 dB) with different numbers of users are calculated 50 times and averaged (each time the users are at different random positions),  as presented in Table \ref{tab2}. Compared to using the 2-D array with the same aperture size, the capacity can be increased by nearly 10$\%$ with 3-D array. Obviously, the benefits brought by the 3-D array in 3GPP channel are smaller than that in Rayleigh channel because of the smaller angular spread at the base station in 3GPP channels. Specifically, the azimuth angle spread at the base station side is regulated to be smaller than 104$^{\circ}$ (angular spread $\theta$ $<$ 52$^{\circ}$) according to 3GPP TR 38.901 \cite{3GPP}. Therefore, when implementing a 3-D array, the specific settings should be carefully taken into consideration, including the array design, multi-path environment, SNR level, user density, etc.
	\begin{table}
		\centering
		%{center}
		\caption{{Benefits of the 3-D array in 3GPP channels}}
		\setlength{\tabcolsep}{5pt}
		\begin{tabular}{|c|c|c|c|}
			\hline
			Scenarios                                                                      & \multicolumn{1}{l|}{2-D array} & \multicolumn{1}{l|}{3-D array} & \multicolumn{1}{l|}{Percent of increase} \\ \hline
			\begin{tabular}[c]{@{}c@{}}Diversity - 100 users\\ (2-D 3GPP UMa)\end{tabular} & 1.83                           & 2.18                           & 16\%                                     \\ \hline
			\begin{tabular}[c]{@{}c@{}}Capacity - 100 users\\ (2-D 3GPP UMa)\end{tabular}  & 18.50                          & 21.08                          & 13\%                                     \\ \hline
			\begin{tabular}[c]{@{}c@{}}Diversity - 500 users\\ (2-D 3GPP UMa)\end{tabular} & 2.56                           & 3.01                           & 17\%                                     \\ \hline
			\begin{tabular}[c]{@{}c@{}}Capacity - 500 users\\ (2-D 3GPP UMa)\end{tabular}  & 21.13                          & 23.80                          & 13\%                                     \\ \hline
			\begin{tabular}[c]{@{}c@{}}Diversity - 100 users\\ (3-D 3GPP UMa)\end{tabular} & 1.65                           & 1.79                           & 8\%                                      \\ \hline
			\begin{tabular}[c]{@{}c@{}}Capacity - 100 users\\ (3-D 3GPP UMa)\end{tabular}  & 16.84                          & 18.17                          & 8\%                                      \\ \hline
			\begin{tabular}[c]{@{}c@{}}Diversity - 500 users\\ (3-D 3GPP UMa)\end{tabular} & 1.80                           & 1.99                           & 11\%                                     \\ \hline
			\begin{tabular}[c]{@{}c@{}}Capacity - 500 users\\ (3-D 3GPP UMa)\end{tabular}  & 17.76                          & 19.02                          & 7\%                                      \\ \hline
		\end{tabular}
		\label{tab2}
		%\end{center}
	\end{table}
	\section{Conclusion}
	In this study, we successfully overcome the fundamental DOF limit in holographic MIMO communications by introducing a 3-D array topology. We systematically elucidate the principles behind the DOF limit and demonstrate the step-by-step utilization of the 3-D array topology to surpass this constraint, progressing from the ideal Clarke model to practical scenarios. Particularly, the benefits of 3-D array are illustrated from the perspectives of electromagnetics and beamforming. The numerical and experimental results further substantiate that, after reaching the DOF limit, the diversities and capacities of MIMO systems can be further improved by applying the 3-D topology. Furthermore, the performances of 3-D arrays in 3GPP channels are investigated, regarding the possible small angular spreads in practical implementations. Consequently, the 3-D array stands out as a promising candidate for bolstering MIMO performance across various scenarios in the future landscape of wireless communications. 
	
	Future endeavors will delve into exploring corresponding channel modeling and precoding technologies. Additionally, we aim to enhance hardware designs, incorporating advanced antenna and decoupling structures to improve radiation patterns and matching. These avenues of research will contribute to the ongoing advancement and optimization of holographic MIMO communication systems.

%%%%%%%%%%%%%%%%%%%%%%%%%%%%%%%%%%%%%%%%%%
\vspace{6pt}

%% Only for journal Encyclopedia
%\entrylink{The Link to this entry published on the encyclopedia platform.}

%\abbreviations{Abbreviations}{
%The following abbreviations are used in this manuscript:\\
%
%\noindent 
%\begin{tabular}{@{}ll}
%MDPI & Multidisciplinary Digital Publishing Institute\\
%DOAJ & Directory of open access journals\\
%TLA & Three letter acronym\\
%LD & Linear dichroism
%\end{tabular}
%}

%%%%%%%%%%%%%%%%%%%%%%%%%%%%%%%%%%%%%%%%%%

	
	%\bibliography{your_external_BibTeX_file}
	
	%=====================================
	% References, variant B: internal bibliography
	%=====================================
	%\bibliographystyle{mdpi}
	\bibliographystyle{unsrt}  
	\bibliography{Bibliography}

\begin{thebibliography}{10}

\bibitem{telatar1999capacity}
Emre Telatar.
\newblock Capacity of multi-antenna {Gaussian} channels.
\newblock {\em Eur. Trans. Telecomm.}, 10(6):585--595, 1999.

\bibitem{tse2005fundamentals}
David Tse and Pramod Viswanath.
\newblock {\em Fundamentals of wireless communication}.
\newblock 2005.

\bibitem{paulraj2004overview}
Arogyaswami~J Paulraj, Dhananjay~A Gore, Rohit~U Nabar, and Helmut Bolcskei.
\newblock An overview of {MIMO} communications-a key to gigabit wireless.
\newblock {\em Proc. IEEE}, 92(2):198--218, 2004.

\bibitem{TL2014}
Erik~G. Larsson, Ove Edfors, Fredrik Tufvesson, and Thomas~L. Marzetta.
\newblock Massive {MIMO} for next generation wireless systems.
\newblock {\em IEEE Commun. Mag.}, 52(2):186--195, 2014.

\bibitem{bjornson2019massive}
Emil Bj{\"o}rnson, Luca Sanguinetti, Henk Wymeersch, Jakob Hoydis, and Thomas~L
  Marzetta.
\newblock Massive {MIMO} is a reality—what is next?: Five promising research
  directions for antenna arrays.
\newblock {\em Digit. Signal Prog.}, 94:3--20, 2019.

\bibitem{Merouane2013}
Jakob Hoydis, Stephan ten Brink, and Merouane Debbah.
\newblock Massive {MIMO} in the {UL/DL} of cellular networks: How many antennas
  do we need?
\newblock {\em IEEE J. Sel. Areas Commun.}, 31(2):160--171, 2013.

\bibitem{Wang2023}
Zhe Wang, Jiayi Zhang, Hongyang Du, Wei E.~I. Sha, Bo~Ai, Dusit Niyato, and
  Merouane Debbah.
\newblock Extremely large-scale {MIMO}: Fundamentals, challenges, solutions,
  and future directions.
\newblock {\em IEEE Wirel. Commun.}, pages 1--9, 2023.

\bibitem{Wu2017}
Ke-Li Wu, Changning Wei, Xide Mei, and Zhen-Yuan Zhang.
\newblock Array-antenna decoupling surface.
\newblock {\em IEEE Trans. Antennas Propag.}, 65(12):6728--6738, 2017.

\bibitem{zhang2019mutual}
Shuai Zhang, Xiaoming Chen, and Gert~Fr{\o}lund Pedersen.
\newblock Mutual coupling suppression with decoupling ground for massive {MIMO}
  antenna arrays.
\newblock {\em IEEE Trans. Veh. Technol.}, 68(8):7273--7282, 2019.

\bibitem{chen2021simultaneous}
Xiaoming Chen, Mengran Zhao, Huilin Huang, Yipeng Wang, Shitao Zhu, Chao Zhang,
  Jianjia Yi, and Ahmed~A Kishk.
\newblock Simultaneous decoupling and decorrelation scheme of {MIMO} arrays.
\newblock {\em IEEE Trans. Veh. Technol.}, 71(2):2164--2169, 2021.

\bibitem{wang2021improvement}
Yipeng Wang, Xiaoming Chen, Xiaobo Liu, Jianjia Yi, Juan Chen, Anxue Zhang, and
  Ahmed~A Kishk.
\newblock Improvement of diversity and capacity of {MIMO} system using
  scatterer array.
\newblock {\em IEEE Trans. Antennas Propag.}, 70(1):789--794, 2021.

\bibitem{cui2020}
Wankai Tang, Jun~Yan Dai, Ming~Zheng Chen, Kai-Kit Wong, Xiao Li, Xinsheng
  Zhao, Shi Jin, Qiang Cheng, and Tie~Jun Cui.
\newblock {MIMO} transmission through reconfigurable intelligent surface:
  System design, analysis, and implementation.
\newblock {\em IEEE J. Sel. Areas Commun.}, 38(11):2683--2699, 2020.

\bibitem{Chongwen2019}
Chongwen Huang, Alessio Zappone, George~C. Alexandropoulos, Mérouane Debbah,
  and Chau Yuen.
\newblock Reconfigurable intelligent surfaces for energy efficiency in wireless
  communication.
\newblock {\em IEEE Trans. Wirel. Commun.}, 18(8):4157--4170, 2019.

\bibitem{wei2022multi}
Li~Wei, Chongwen Huang, George~C Alexandropoulos, Wei E.~I. Sha, Zhaoyang
  Zhang, M{\'e}rouane Debbah, and Chau Yuen.
\newblock Multi-user holographic {MIMO} surfaces: Channel modeling and spectral
  efficiency analysis.
\newblock {\em IEEE J. Sel. Top. Signal Process.}, 16(5):1112--1124, 2022.

\bibitem{An2023}
Jiancheng An, Chao Xu, Derrick Wing~Kwan Ng, George~C. Alexandropoulos,
  Chongwen Huang, Chau Yuen, and Lajos Hanzo.
\newblock Stacked intelligent metasurfaces for efficient holographic {MIMO}
  communications in 6g.
\newblock {\em IEEE J. Sel. Areas Commun.}, 41(8):2380--2396, 2023.

\bibitem{Wei2023}
Li~Wei, Chongwen Huang, George~C. Alexandropoulos, Zhaohui Yang, Jun Yang, Wei
  E.~I. Sha, Zhaoyang Zhang, Mérouane Debbah, and Chau Yuen.
\newblock Tri-polarized holographic {MIMO} surfaces for near-field
  communications: Channel modeling and precoding design.
\newblock {\em IEEE Trans. Wirel. Commun.}, pages 1--1, 2023.

\bibitem{ShuaiOJAP}
Shuai S.~A. Yuan, Xiaoming Chen, Chongwen Huang, and Wei E.~I. Sha.
\newblock Effects of mutual coupling on degree of freedom and antenna
  efficiency in holographic {MIMO} communications.
\newblock {\em IEEE Open J. Antennas Propag.}, 4:237--244, 2023.

\bibitem{Zhang2023}
Yuan Zhang, Jianhua Zhang, Yuxiang Zhang, Yuan Yao, and Guangyi Liu.
\newblock Capacity analysis of holographic {MIMO} channels with practical
  constraints.
\newblock {\em IEEE Wireless Commun. Lett.}, 12(6):1101--1105, 2023.

\bibitem{Demi2022}
Ozlem~Tugfe Demir, Emil Bjornson, and Luca Sanguinetti.
\newblock Channel modeling and channel estimation for holographic massive
  {MIMO} with planar arrays.
\newblock {\em IEEE Wireless Commun. Lett.}, 11(5):997--1001, 2022.

\bibitem{Gong2023}
Tierui Gong, Panagiotis Gavriilidis, Ran Ji, Chongwen Huang, George~C.
  Alexandropoulos, Li~Wei, Zhaoyang Zhang, Mérouane Debbah, H.~Vincent Poor,
  and Chau Yuen.
\newblock Holographic {MIMO} communications: Theoretical foundations, enabling
  technologies, and future directions.
\newblock {\em IEEE Commun. Surv. Tutor.}, early access, 2023. DOI:
  10.1109/COMST.2023.3309529.

\bibitem{gong2023holographic}
Tierui Gong, Li~Wei, Chongwen Huang, Zhijia Yang, Jiguang He, M{\'e}rouane
  Debbah, and Chau Yuen.
\newblock Holographic {MIMO} communications with arbitrary surface placements:
  Near-field {LoS} channel model and capacity limit.
\newblock {\em IEEE J. Sel. Areas Commun.}, accepted, 2023. [Online].
  Available: https://arxiv.org/abs/2304.05259.

\bibitem{gong2023transmit}
Tierui Gong, Chongwen Huang, Jiguang He, Marco Di~Renzo, M{\'e}rouane Debbah,
  and Chau Yuen.
\newblock A transmit-receive parameter separable electromagnetic channel model
  for {LoS} holographic {MIMO}.
\newblock In {\em Proc. 2023 IEEE Glob. Commun. Conf. (GLOBECOM), IEEE}, pages
  5707--5712, Dec. 2023.

\bibitem{Pizzo2020}
Andrea Pizzo, Thomas~L. Marzetta, and Luca Sanguinetti.
\newblock Spatially-stationary model for holographic {MIMO} small-scale fading.
\newblock {\em IEEE J. Sel. Areas Commun.}, 38(9):1964--1979, 2020.

\bibitem{an2023tutorial1}
Jiancheng An, Chau Yuen, Chongwen Huang, M{\'e}rouane Debbah, H~Vincent Poor,
  and Lajos Hanzo.
\newblock A tutorial on holographic {MIMO} communications—part i: Channel
  modeling and channel estimation.
\newblock {\em IEEE Commun. Lett.}, 2023.

\bibitem{an2023tutorial2}
Jiancheng An, Chau Yuen, Chongwen Huang, M{\'e}rouane Debbah, H~Vincent Poor,
  and Lajos Hanzo.
\newblock A tutorial on holographic {MIMO} communications—part ii:
  Performance analysis and holographic beamforming.
\newblock {\em IEEE Commun. Lett.}, 2023.

\bibitem{an2023tutorial3}
Jiancheng An, Chau Yuen, Chongwen Huang, M{\'e}rouane Debbah, H~Vincent Poor,
  and Lajos Hanzo.
\newblock A tutorial on holographic {MIMO} communications—part iii: Open
  opportunities and challenges.
\newblock {\em IEEE Commun. Lett.}, 2023.

\bibitem{Huang2020}
Chongwen Huang, Sha Hu, George~C. Alexandropoulos, Alessio Zappone, Chau Yuen,
  Rui Zhang, Marco~Di Renzo, and Merouane Debbah.
\newblock Holographic {MIMO} surfaces for 6g wireless networks: Opportunities,
  challenges, and trends.
\newblock {\em IEEE Wirel. Commun.}, 27(5):118--125, 2020.

\bibitem{Thomas2020}
Andrea Pizzo, Thomas~L. Marzetta, and Luca Sanguinetti.
\newblock Degrees of freedom of holographic {MIMO} channels.
\newblock In {\em 2020 IEEE 21st International Workshop on Signal Processing
  Advances in Wireless Communications (SPAWC)}, pages 1--5, 2020.

\bibitem{MD2019}
Marco~Donald Migliore.
\newblock Horse (electromagnetics) is more important than horseman
  (information) for wireless transmission.
\newblock {\em IEEE Trans. Antennas Propag.}, 67(4):2046--2055, 2019.

\bibitem{Shuai2021}
Shuai S.~A. Yuan, Zi~He, Xiaoming Chen, Chongwen Huang, and Wei E.~I. Sha.
\newblock Electromagnetic effective degree of freedom of an {MIMO} system in
  free space.
\newblock {\em IEEE Antennas Wirel. Propag. Lett.}, 21(3):446--450, 2022.

\bibitem{Dai2023}
Zhongzhichao Wan, Jieao Zhu, Zijian Zhang, Linglong Dai, and Chan-Byoung Chae.
\newblock Mutual information for electromagnetic information theory based on
  random fields.
\newblock {\em IEEE Trans Commun.}, 71(4):1982--1996, 2023.

\bibitem{Han2023}
Zixiang Han, Shanpu Shen, Yujie Zhang, Shiwen Tang, Chi-Yuk Chiu, and Ross
  Murch.
\newblock Using loaded n-port structures to achieve the continuous-space
  electromagnetic channel capacity bound.
\newblock {\em IEEE Trans. Wirel. Commun.}, pages 1--1, 2023.

\bibitem{Jeon2018}
Wonseok Jeon and Sae-Young Chung.
\newblock Capacity of continuous-space electromagnetic channels with lossy
  transceivers.
\newblock {\em IEEE Trans. Inf. Theory}, 64(3):1977--1991, 2018.

\bibitem{franceschetti_2017}
Massimo Franceschetti.
\newblock {\em Wave Theory of Information}.
\newblock Cambridge University Press, 2017.

\bibitem{gustafsson2023degrees}
Mats Gustafsson and Johan Lundgren.
\newblock Degrees of freedom and characteristic modes.
\newblock 2023.

\bibitem{Dardari2021}
Davide Dardari and Nicolò Decarli.
\newblock Holographic communication using intelligent surfaces.
\newblock {\em IEEE Commun. Mag.}, 59(6):35--41, 2021.

\bibitem{Li2023}
Ruifeng Li, Da~Li, Jinyan Ma, Zhaoyang Feng, Ling Zhang, Shurun Tan, Wei E.~I.
  Sha, Hongsheng Chen, and Er-Ping Li.
\newblock An electromagnetic information theory based model for efficient
  characterization of {MIMO} systems in complex space.
\newblock {\em IEEE Trans. Antennas Propag.}, 71(4):3497--3508, 2023.

\bibitem{bai2024information}
Xuyang Bai, Shurun Tan, Said Mikki, Erping Li, and Tie-Jun Cui.
\newblock Information-theoretic measures for reconfigurable metasurface-enabled
  direct digital modulation systems: An electromagnetic perspective.
\newblock {\em Prog. Electromagn. Res.}, 179:1--18, 2024.

\bibitem{balanis2015antenna}
Constantine~A Balanis.
\newblock {\em Antenna theory: analysis and design}.
\newblock 2015.

\bibitem{Hannan1964}
P.~Hannan.
\newblock The element-gain paradox for a phased-array antenna.
\newblock {\em IEEE Trans. Antennas Propag.}, 12(4):423--433, 1964.

\bibitem{Bucci1989}
O.M. Bucci and G.~Franceschetti.
\newblock On the degrees of freedom of scattered fields.
\newblock {\em IEEE Trans. Antennas Propag.}, 37(7):918--926, 1989.

\bibitem{ShuaiPra}
Shuai S.~A. Yuan, Jie Wu, Menglin L.~N. Chen, Zhihao Lan, Liang Zhang, Sheng
  Sun, Zhixiang Huang, Xiaoming Chen, Shilie Zheng, Li~Jun Jiang, Xianmin
  Zhang, and Wei E.~I. Sha.
\newblock Approaching the fundamental limit of orbital-angular-momentum
  multiplexing through a hologram metasurface.
\newblock {\em Phys. Rev. Applied}, 16:064042, 2021.

\bibitem{Mats2021}
Casimir Ehrenborg, Mats Gustafsson, and Miloslav Capek.
\newblock Capacity bounds and degrees of freedom for {MIMO} antennas
  constrained by {Q}-factor.
\newblock {\em IEEE Trans. Antennas Propag.}, 69(9):5388--5400, 2021.

\bibitem{Wallace2008}
Michael~A. Jensen and Jon~W. Wallace.
\newblock Capacity of the continuous-space electromagnetic channel.
\newblock {\em IEEE Trans. Antennas Propag.}, 56(2):524--531, 2008.

\bibitem{signal2005space}
A.S.Y. Poon, R.W. Brodersen, and D.N.C. Tse.
\newblock Degrees of freedom in multiple-antenna channels: a signal space
  approach.
\newblock {\em IEEE Trans. Inf. Theory}, 51(2):523--536, 2005.

\bibitem{Shuai2022}
Shuai S.~A. Yuan, Zi~He, Sheng Sun, Xiaoming Chen, Chongwen Huang, and Wei
  E.~I. Sha.
\newblock Electromagnetic effective-degree-of-freedom limit of a {MIMO} system
  in {2-D} inhomogeneous environment.
\newblock {\em Electronics}, 11(19):3232, 2022.

\bibitem{Shen2023}
Yinsong Shen, Zi~He, Wei E.~I. Sha, Shuai S.~A. Yuan, and Xiaoming Chen.
\newblock Electromagnetic effective-degree-of-freedom prediction with parabolic
  equation method.
\newblock {\em IEEE Trans. Antennas Propag.}, 71(4):3752--3757, 2023.

\bibitem{piestun2000electromagnetic}
Rafael Piestun and David~AB Miller.
\newblock Electromagnetic degrees of freedom of an optical system.
\newblock {\em J. Opt. Soc. Am. A-Opt. Image Sci. Vis.}, 17(5):892--902, 2000.

\bibitem{miller2000communicating}
David~AB Miller.
\newblock Communicating with waves between volumes: evaluating orthogonal
  spatial channels and limits on coupling strengths.
\newblock {\em Appl. Optics}, 39(11):1681--1699, 2000.

\bibitem{hu2018beyond}
Sha Hu, Fredrik Rusek, and Ove Edfors.
\newblock Beyond massive {MIMO}: The potential of data transmission with large
  intelligent surfaces.
\newblock {\em IEEE Trans. Signal Process.}, 66(10):2746--2758, 2018.

\bibitem{Song2015}
Xiaohang Song and Gerhard Fettweis.
\newblock On spatial multiplexing of strong line-of-sight {MIMO} with {3D}
  antenna arrangements.
\newblock {\em IEEE Wirel. Commun. Lett.}, 4(4):393--396, 2015.

\bibitem{an2023stacked}
Jiancheng An, Chau Yuen, Chao Xu, Hongbin Li, Derrick Wing~Kwan Ng, Marco
  Di~Renzo, M{\'e}rouane Debbah, and Lajos Hanzo.
\newblock Stacked intelligent metasurface-aided {MIMO} transceiver design.
\newblock {\em arXiv preprint arXiv:2311.09814}, 2023.

\bibitem{an2023stacked2}
Jiancheng An, Chao Xu, Derrick Wing~Kwan Ng, George~C Alexandropoulos, Chongwen
  Huang, Chau Yuen, and Lajos Hanzo.
\newblock Stacked intelligent metasurfaces for efficient holographic {MIMO}
  communications in 6g.
\newblock {\em IEEE J. Sel. Areas Commun.}, 2023.

\bibitem{clarke1968statistical}
Richard~Hedley Clarke.
\newblock A statistical theory of mobile-radio reception.
\newblock {\em Bell Syst. Tech. J.}, 47(6):957--1000, 1968.

\bibitem{clarke1997}
R.~H. Clarke and Wee~Lin Khoo.
\newblock {3-D} mobile radio channel statistics.
\newblock {\em IEEE Trans. Veh. Technol.}, 46(3):798--799, 1997.

\bibitem{Andersen2002}
J.B. Andersen and K.I. Pedersen.
\newblock Angle-of-arrival statistics for low resolution antennas.
\newblock {\em IEEE Trans. Antennas Propag.}, 50(3):391--395, 2002.

\bibitem{miller2019waves}
David~AB Miller.
\newblock Waves, modes, communications, and optics: a tutorial.
\newblock {\em Adv. Opt. Photonics}, 11(3):679--825, 2019.

\bibitem{NJ2005}
M.T. Ivrlac and J.A. Nossek.
\newblock Diversity and correlation in {Rayleigh} fading {MIMO} channels.
\newblock In {\em 2005 IEEE 61st Vehicular Technology Conference}, volume~1,
  pages 151--155 Vol. 1, 2005.

\bibitem{Verdu2002}
S.~Verdu.
\newblock Spectral efficiency in the wideband regime.
\newblock {\em IEEE Trans. Inf. Theory}, 48(6):1319--1343, 2002.

\bibitem{xiaoming2013}
Xiaoming Chen, Per-Simon Kildal, Jan Carlsson, and Jian Yang.
\newblock {MRC} diversity and {MIMO} capacity evaluations of multi-port
  antennas using reverberation chamber and anechoic chamber.
\newblock {\em IEEE Trans. Antennas Propag.}, 61(2):917--926, 2013.

\bibitem{Loyka2009}
Sergey Loyka and Georgy Levin.
\newblock On physically-based normalization of {MIMO} channel matrices.
\newblock {\em IEEE Trans. Wirel. Commun.}, 8(3):1107--1112, 2009.

\bibitem{Kildal2004}
P.-S. Kildal and K.~Rosengren.
\newblock Correlation and capacity of {MIMO} systems and mutual coupling,
  radiation efficiency, and diversity gain of their antennas: simulations and
  measurements in a reverberation chamber.
\newblock {\em IEEE Commun. Mag.}, 42(12):104--112, 2004.

\bibitem{xiaoming2017}
Xiaoming Chen and Shuai Zhang.
\newblock Multiplexing efficiency for {MIMO} antenna-channel impairment
  characterisation in realistic multipath environments.
\newblock {\em IET Microw. Antennas Propag.}, 11:524--528(4), 2017.

\bibitem{Kildal2016}
Per-Simon Kildal, Abbas Vosoogh, and Stefano Maci.
\newblock Fundamental directivity limitations of dense array antennas: A
  numerical study using {H}annan’s embedded element efficiency.
\newblock {\em IEEE Antennas Wirel. Propag. Lett.}, 15:766--769, 2016.

\bibitem{kildal2015foundations}
Per-Simon Kildal.
\newblock {\em Foundations of antenna engineering: a unified approach for
  line-of-sight and multipath}.
\newblock Gothenburg, Sweden, 2015.

\bibitem{yang2009electromagnetic}
Fan Yang and Yahya Rahmat-Samii.
\newblock {\em Electromagnetic band gap structures in antenna engineering}.
\newblock Cambridge university press Cambridge, UK, 2009.

\bibitem{abedin2005effects}
M~Faisal Abedin and Mohammod Ali.
\newblock Effects of {EBG} reflection phase profiles on the input impedance and
  bandwidth of ultrathin directional dipoles.
\newblock {\em IEEE Trans. Antennas Propag.}, 53(11):3664--3672, 2005.

\bibitem{yang2003reflection}
Fan Yang and Yahya Rahmat-Samii.
\newblock Reflection phase characterizations of the {EBG} ground plane for low
  profile wire antenna applications.
\newblock {\em IEEE Trans. Antennas Propag.}, 51(10):2691--2703, 2003.

\bibitem{Quadriga}
Stephan Jaeckel, Leszek Raschkowski, Kai Börner, and Lars Thiele.
\newblock {QuaDRiGa}: A 3-{D} multi-cell channel model with time evolution for
  enabling virtual field trials.
\newblock {\em IEEE Trans. Antennas Propag.}, 62(6):3242--3256, 2014.

\bibitem{3GPP}
Study on channel model for frequencies from 0.5 to 100 {GHz}, version
  16.1.0,” {3GPP, Sophia Antipolis, France, 3GPP Rep. (TR) 38.901, 2020.}
  [online]. available: http://www.3gpp.org/dynareport/38901.htm.

\end{thebibliography}
\end{document}